\documentclass[12pt]{article}
\usepackage{jheppub}
\usepackage{amssymb,amsfonts}

\usepackage{natbib}
\setcitestyle{square, comma, numbers, compress}
\usepackage{wrapfig}
\usepackage{epsfig}
\usepackage[T1]{fontenc}
\usepackage[utf8]{inputenc}
\usepackage{lmodern}
\usepackage{float}
\usepackage{placeins}
\usepackage{dsfont}
\usepackage{arydshln}
\usepackage{caption}

\usepackage[dvipsnames]{xcolor}
\usepackage{tikz}

\def\be{\begin{eqnarray}}
\def\ee{\end{eqnarray}}
\newcommand{\nn}{\nonumber}
\newcommand\para{\paragraph{}}
\newcommand{\ft}[2]{{\textstyle\frac{#1}{#2}}}
\newcommand{\eqn}[1]{(\ref{#1})}

\newcommand{\ra}{\rightarrow}

\def\Dslash{\,\,{\raise.15ex\hbox{/}\mkern-12mu D}}
\def\Dbarslash{\,\,{\raise.15ex\hbox{/}\mkern-12mu {\bar D}}}
\def\delslash{\,\,{\raise.15ex\hbox{/}\mkern-9mu \partial}}
\def\delbarslash{\,\,{\raise.15ex\hbox{/}\mkern-9mu {\bar\partial}}}
\def\pslash{\,\,{\raise.15ex\hbox{/}\mkern-9mu p}}
\def\calDslash{\,\,{\raise.15ex\hbox{/}\mkern-12mu {\cal D}}}

\newcommand{\D}{{\cal D}}

\newcommand{\vev}{VEV }

\newcommand{\Z}{{\mathds Z}}

\def\calL{\mathcal{L}}

\def\calO{\mathcal{O}}

\def\lae{\mathrel{\mathop{\smash{\lower .5 ex \hbox{$\stackrel<\sim$}}}}}
\def\lae{\mathrel{\mathop{\smash{\lower .5 ex \hbox{$\stackrel>\sim$}}}}}

\newcommand{\tvp}{\tilde{\varphi}}

\title{Phases of 2d Gauge Theories and Symmetric Mass Generation}

\author[a]{Rishi Mouland,}
\author[b]{David Tong,}
\author[c]{and Bernardo Zan}

\affiliation[a]{Abdus Salam Centre for Theoretical Physics, Imperial College London, London SW7 2AZ, UK}
\affiliation[b]{Department of Applied Mathematics and Theoretical Physics \\ University Cambridge, CB3 0WA, UK}
\affiliation[c]{Dipartimento di Fisica, Università di Genova and INFN, Sezione di Genova, Via
Dodecaneso 33, 16146, Genoa, Italy}

\abstract{We study the dynamics and phase structure of Abelian gauge theories in $d=1+1$ dimensions. These include $U(1)$ gauge theory coupled to a scalar and a fermion, as well as the two-flavour Schwinger model with different charges. Both theories exhibit a surprisingly rich phase diagram as  masses are varied, with both  $c=1$ and $c=1/2$ critical lines or points. We build up to the study of 2d chiral gauge theories, which hold particular interest because they provide a mechanism for symmetric mass generation, a phenomenon in which fermions become gapped without breaking chiral symmetries. 
}

\begin{document}
\pagestyle{plain} \setcounter{page}{1}
\newcounter{bean}
\baselineskip16pt \setcounter{section}{0}

\maketitle


\section{Introduction}

The purpose of this paper is to address some basic issues about the dynamics and phase structure of    2d Abelian gauge theories.

\para
Our motivation for exploring these theories is a phenomenon known as {\it symmetric mass generation}. This is a mechanism whereby fermions become gapped without breaking chiral (but non-anomalous) global symmetries. These chiral symmetries would be explicitly broken by a quadratic mass term for the fermions, but need not be broken if the fermions get a mass through some strong coupling mechanism. 

\para
As we explain later in this introduction, one way of implementing symmetric mass generation is through chiral gauge dynamics. The ultimate goal of this paper is to explore a  class of Abelian chiral gauge theories in some detail to better understand this mechanism. However, along the way we will have need to study a number of other simple gauge theories, both chiral and non-chiral, and determine their phase structure.

\para
There is a standard tool to solve 2d Abelian gauge theories: bosonisation. Done properly, bosonisation is fiddly. A compact boson is not the same as a fermion in two dimensions. Instead, the  equivalence holds only after implementing a $\Z_2$ gauging. For many applications, such as solving the Schwinger model through bosonisation \cite{schwinger,cjackiw,coleman}, this subtlety can be largely ignored. But when we come to study chiral gauge theories through bosonisation, such  subtleties are important and must be treated carefully to determine, for example, the number of ground states of the theory. Part of this paper -- the fiddly part -- is devoted to getting these things right. 

\para
In the rest of this introduction, we describe some of the questions that we are interested in, and the theories that we explore. We also  summarise our results.

\subsubsection*{QED With a Scalar and a Fermion}

The 2d Abelian Higgs model has Lagrangian 
\be {\cal L} = \frac{1}{2e^2} F_{01}^2 + |{\cal D}_\mu\phi|^2 - m_s^2|\phi|^2 - \frac{\lambda}{2}|\phi|^4\ .\label{ah}\ee
When $m_s^2\gg e^2$, the scalar can be safely integrated out and we're left with 2d Maxwell theory. This theory confines any external charges, with Wilson loops exhibiting an area law. When $m_s^2\ll -e^2$,  the theory sits in the Higgs phase and one might naively expect a  perimeter law for the Wilson loop. However, it was appreciated long ago that this classical intuition misses the effect of vortices which, in 2d, play the role of instantons \cite{dashen,aspects}. After summing over the vortices, the Wilson loop once again exhibits an area law, albeit with an exponentially reduced string tension.

\para
The first question that we want to ask is: what happens if we add massless, charged fermions to the theory in the Higgs phase?  This question was addressed long ago \cite{grossy} and, roughly speaking, the answer is as follows: if we add $N_f$ massless Dirac fermions, then the theory is gapless for $N_f>1$ and gapped for $N_f=1$. This follows largely on symmetry grounds. For $N_f>1$, there is an $SU(N_f)_L\times SU(N_f)_R$ chiral symmetry  whose current algebra guarantees gapless modes. However, for $N_f=1$, the ABJ anomaly means that there is no additional chiral symmetry and nothing to prohibit the theory from becoming gapped.

\para
Here we revisit the theory with $N_f=1$. Specifically, we want to understand the zero-temperature phase diagram of this theory as we vary the scalar mass $m_s^2$ and the fermion mass $m_f$. Despite the simplicity of the theory, this phase diagram has not, to our knowledge, been previously constructed.  

\para
In Section \ref{scalarfermionsec}, we will argue that, as we vary the fermion mass $m_f$  there is a value at which a gapless mode emerges in the Higgs phase. This result is similar in spirit to the fact that, for 4d QCD with  a single $N_f=1$ fermion, there is a value of the fermion mass for which the $\eta'$ meson becomes massless \cite{seiberg}. In 2d, we will see that we  have a critical line, rather than a critical point, which extends in the $(m_s^2,m_f)$-plane.  
This is a line of $c=1$ CFTs which terminates at a point where it splits into two $c=\ft 12$ lines. Our  proposal for the phase diagram is sketched in Figure \ref{phase2fig}. 

\subsubsection*{QED With Two Dirac Fermions}

Consider $U(1)$ gauge theory coupled to two Dirac fermions. For the case where the two fermions both have charge $q=1$, the phase diagram was analysed only recently and exhibits a surprisingly rich structure  as the masses are varied \cite{bernardo}.   We  reproduce this phase diagram in Figure \ref{phaseqed2fig}.

\para
When the two fermions are massless, the result is less surprising: the theory flows to a $c=1$ fixed point with an $SU(2)$ global symmetry. In Section \ref{twofermionsec}, we extend this result to the case where the fermions have co-prime charges $p$ and $q$. We show that the theory with massless fermions flows to a $c=1$ fixed point, described by a compact boson with radius $R$ given by
\be R^2 = \frac{p^2+q^2}{2}\ .\label{r2}\ee
Here we use the convention where the self-dual radius, exhibiting enhanced $SU(2)$ symmetry, is at $R=1$.

\para
This is the first place that the $\Z_2$ subtleties involved in bosonisation rear their head, albeit in a mild way. We should distinguish between the situation where $p$ and $q$ are both odd, and where one of them is even. In the former case, the gauge theory is bosonic, meaning that $(-1)^F$ is part of the $U(1)$ gauge group and all gauge-invariant operators are Grassmann-even. In this case, the infra-red limit is  indeed  the compact boson  with radius \eqn{r2}.

\para
However, when one of $p$ or $q$ is even, the gauge theory is fermionic, meaning that there are gauge-invariant Grassmann-odd operators. Correspondingly, the low-energy theory should be a fermionic CFT that    is sensitive to the spin structure. This fermionic CFT can be viewed as a compact boson of radius \eqn{r2}, coupled to $\Z_2$ gauge field that gauges part of the winding symmetry. 

\para
We also analyse the phase structure of the two-flavour Schwinger model with charges $p$ and $q$ as the fermion masses are varied. This too depends on whether the product $pq$ is odd or even, on whether $p$ or $q$ is equal to 1, and, to a lesser extent, on whether $p$ or $q$ is equal to 2.  The resulting phase diagrams are shown in Figures \ref{phaseqed22fig} and \ref{phaseqed33fig} for $pq$ odd and even, respectively. 

\subsubsection*{Chiral QED}

In Section \ref{chiralsec}, we turn to chiral gauge theories. The simplest chiral gauge theory consists of a $U(1)$ gauge field coupled to two left-moving fermions with charges 3 and 4, and two right-moving fermions with charges 5 and 0. The anomalies cancel by virtue of the fact
\be 3^2 + 4^2 = 5^2 + 0^2\ .\ee
We would like to know the infra-red behaviour of this theory. 

\para
At first sight, the answer seems obvious. The right-mover with charge 0 is clearly a spectator in the RG flow and anomaly matching means that, when the dust settles, the gauge sector must confine to give a single left-moving massless fermion. At low-energies, the theory should therefore be described by a single gapless Dirac fermion.

\para
The problem with this argument is that it doesn't tell us anything about potential TQFTs that might accompany the low-energy dynamics. Might it be possible that the low-energy physics comprises of a massless Dirac fermion tensored with a TQFT? In Section \ref{chiralsec}, we show that for the 3450 model (but not necessarily for other chiral gauge theories), the answer is no: the low-energy dynamics is what you naively guess: a single Dirac fermion\footnote{This result was previously derived in unpublished work by Philip Boyle Smith in 2024, using a direct Hilbert space method.  We thank Philip for sharing his results with us.}.

\subsubsection*{Symmetric Mass Generation}

As mentioned above, our motivation for revisiting these old question about 2d gauge theories comes from the topic of symmetric mass generation. (See \cite{review} for a review.) 

\para
A rather simple model of symmetric mass generation  was proposed in  \cite{mesmg}, involving gauge fields coupled to both chiral fermions and scalars; the details of this model can be found in Section \ref{smgsec}.  The idea is that the Higgs phase of the theory has massless fermions, protected by a chiral global symmetry, while the confining phase is gapped and the global symmetry is unbroken. This behaviour contrasts with  the QED-with-scalar-and-fermion story that we sketched above: in that case any gapless modes are necessarily fine-tuned and, while there are lines of critical points, there is no gapless phase that is immune to all relevant perturbations. But, as we show,  chiral theories are different. 

\para
In Section \ref{smgsec}, we determine the phase diagram of this class of chiral gauge theories. We will see that the Higgs phase of the theory is  indeed gapless, with no relevant operators consistent with the symmetries of the theory. But as we vary the scalar masses, and hence the scalar vacuum expectation values (VEVs), the dimensions of operators changes. At some point we move into the confining phase of the theory, gapping all fermions without breaking the global chiral symmetry.

\section{QED with a Scalar and a Fermion}\label{scalarfermionsec}

In this section, we study $U(1)$ gauge theory with a single complex scalar $\phi$ and a Dirac fermion $\psi$. The Lagrangian is 
\be   {\cal L} = \frac{1}{2e^2} F_{01}^2 + |{\cal D}_\mu\phi|^2 - m_s^2|\phi|^2 - \frac{\lambda}{2}|\phi|^4 + i\bar{\psi} \Dslash \psi - im_f \bar{\psi}\psi\ .\label{1and1}\ee 
Both $\phi$ and $\psi$ have charge 1 under the gauge symmetry. We set the theta angle to $\theta=0$. (This is where the phase diagram is richest; we comment on the $\theta\neq 0$ case below.) 

\para
The theory has a single $U(1)_V$ flavour symmetry which can be taken to act on the fermion as
\be U(1)_V: \psi \ra e^{-i\alpha}\psi\label{u1v}\ ,\ee
while leaving the scalar untouched. 
When $m_f=0$, the classical theory has an additional axial symmetry $U(1)_A: \psi \ra e^{-i\alpha \gamma^3}\psi$, but this suffers the usual chiral anomaly and is not a symmetry of the quantum theory. Part of the surprise in this theory is that we'll see the anomalous $U(1)_A$ symmetry emerging as an accidental symmetry in some parts of the phase diagram.

\para
In $d=1+1$, the scalar is dimensionless. This means that there are an infinite number of relevant operators that we could add to the action, including $|\phi|^{2n}$ and Yukawa terms of the form $|\phi|^{2n} \bar{\psi} \psi$, both of which are invariant under all the symmetries.  We set these operators to zero in the UV; they will not be dynamically generated in the asymptotic (i.e. large mass) region of  the phase diagram. 

\para
We would like to understand the phases of this theory as we vary the fermion mass $m_f$ and the scalar  mass $m_s^2$. When one or the other of these masses is large, we reduce to  the Abelian Higgs model or to the Schwinger model, respectively. For that reason, it will be useful to review the dynamics of these well-studied theories. We look at each in turn before offering a proposal for the full phase diagram in the $(m_f,m_s^2)$-plane in Section \ref{sfsec}.

\subsection{The Abelian-Higgs and Schwinger Models}\label{qedsec}

When  $|m_f|\gg e$, we can integrate out the fermion to be left with the Abelian Higgs model
\be {\cal L} = \frac{1}{2e^2}F_{01}^2 + \frac{\theta}{2\pi} F_{01} + |{\cal D}_\mu\phi|^2 -m_s^2 |\phi|^2- \lambda |\phi|^4 \ .\ee
For $m_f>0$, we have $\theta=0$ while,  for $m_f<0$, integrating out the fermion generates a theta angle  $\theta=\pi$. 

\para
The dynamics of this theory were understood long ago \cite{dashen,aspects} and a full analysis of the phase structure can be found in  \cite{zohart}. Here we review the essential features:

\subsubsection*{Confining Phase: $m_s^2\gg e^2$}

When the scalar is very massive, we may integrate it out and we're left just with pure Maxwell theory. Although there are no propagating degrees of freedom at low energy, there's still some interesting physics. To see this, it's best to work in the $A_0 = 0$ gauge, put the theory on a spatial circle ${\bf S}^1$ of radius $L$, and look at the Wilson line
\be \alpha(t) = \int_0^{2\pi L} dx\ A_1(x,t)\ .\ee
Large gauge transformations imply that this scalar is periodic, with $\alpha \in [0,2\pi)$. The dynamics of pure Maxwell theory is then governed by the quantum mechanical Lagrangian
\be S = \int dt\ \left(\frac{1}{4\pi e^2L} \dot{\alpha}^2 + \frac{\theta}{2\pi} \dot{\alpha}\right)\ .\ee
This takes the same form as for a   particle moving around a solenoid. The theta term doesn't affect the classical equations of motion, but it does affect the spectrum of the quantum theory through its appearance in the canonical momentum $p$. The Hamiltonian is given by
\be H = \frac{1}{4\pi e^2 L} \dot{\alpha}^2 = \pi e^2 L\left(p-\frac{\theta}{2\pi}\right)^2 \ \ \ {\rm with}\ \ \ p= \frac{1}{2\pi e^2 L}\dot{\alpha} + \frac{\theta}{2\pi}\ .\ee
The eigenvalues of $p$ are quantised and the energy eigenvalues can be read off directly from $H$ simply by replacing the operator $p$ with an integer  $p\in \Z$.

\para
Something special  happens as we vary $\theta$. For most values of $\theta$, there is a  unique ground state. However when $\theta = \pi$, the ground state is degenerate, with $p=0$ and $p=1$ both giving the same energy. (A similar statement holds when $\theta = (2n+1)\pi$, with the $p=n$ and $p=n+1$ states degenerate.) This can be traced to a mixed anomaly between time reversal and shifts of $\alpha$, as explained in \cite{ttt}.

\para
The states in pure Maxwell theory are labelled by $p \in \Z$. These can be thought of as states of different, constant electric fields
\be F_{01} = \frac{1}{2\pi L}\dot{\alpha} = e^2\left(p -\frac{\theta}{2\pi}\right)\ .\ee
In particular, the role of the theta term is to turn on a background electric field.

\para
In pure Maxwell theory, all states labelled by $p\in \Z$ are stable energy eigenstates. This is no longer true when we take into account the dynamical, but massive, scalar field $\phi$. For general values of $\theta$, only the ground state $n=0$ is stable. For $\theta=\pi$, the two degenerate ground states are both stable. (As explained in \cite{zohart}, this story is richer if the scalar field is taken to have charge $q>1$.)

\subsubsection*{Higgs Phase: $m_s^2\ll -e^2$}

For $m^2_s$ large and negative, the scalar condenses and we sit in the Higgs phase.  We write
\be  \phi = v e^{i\sigma}\ .\ee
If we ignore the radial mode then we're left with a theory for the periodic scalar $\sigma\in [0,2\pi)$ and Lagrangian
\be {\cal L} = \frac{1}{2e^2} F_{01}^2 + \frac{\theta}{2\pi} F_{01} + v^2 \left(\partial_\mu \sigma + A_\mu\right)^2 \ .\label{scalar}\ee
This has a unique ground state. (Again, the story is richer if $\phi$ has charge $q>1$.)

\para
Crucially, the theory has a unique ground state regardless of the value of $\theta$. That coincides with the story in the confining phase for $\theta\neq \pi$ which means that we expect to   move smoothly from confining to Higgs phases when $\theta\neq \pi$. However, there is also a single ground state in the Higgs phase when $\theta=\pi$, contrasting with the two ground states in the confining phase. This means that there must be a phase transition as we vary the scalar mass from $m^2_s\ll -e^2$ to $m^2_s \gg e^2$ at $\theta=\pi$. The simplest possibility is that the two ground states merge into a single ground state at an Ising point.

\subsubsection*{The Schwinger model}

The story above holds when the fermion mass $|m_f|\gg e$. We can also make progress when the scalar mass  $m_s^2\gg e^2$. In this case, we can integrate out the scalar field  to be left with QED$_2$, also known as the Schwinger model. This has been well studied starting in \cite{schwinger,cjackiw,coleman}.

\para
For generic values of the mass $m_f$, the  Schwinger model is gapped. However, it is known to exhibit a second order phase transition as  we vary the fermion mass from $m_f\ll -e$ to $m_f\gg e$ \cite{shankar}.  This follows from the arguments above: when $m_f\gg e$, we are left with pure Maxwell theory with $\theta=0$ and, correspondingly, a single ground state, while when $m_f\ll -e$, we have pure Maxwell theory with $\theta=\pi$ and two ground states. Numerical studies strongly suggest that the second order transition is an Ising transition and occurs at $m\approx -e/3$ \cite{numerics1,numerics2} (see \cite{ArguelloCruz:2024xzi} for the current most precise determination building on results of \cite{Dempsey:2022nys}.)

\subsubsection*{A First Look at the Phase Diagram}

We now return to the theory \eqn{1and1} with  both a scalar $\phi$ and a  fermion $\psi$.  What does the phase diagram look like as we vary the masses, $m_s^2$ for the scalar and $m_f$ for the fermion? As we now explain, from the semi-classical limits described above, we can identify three different lines of second order phase transitions.

\begin{figure}[htb]
\begin{center}
\epsfxsize=2.5in\leavevmode\epsfbox{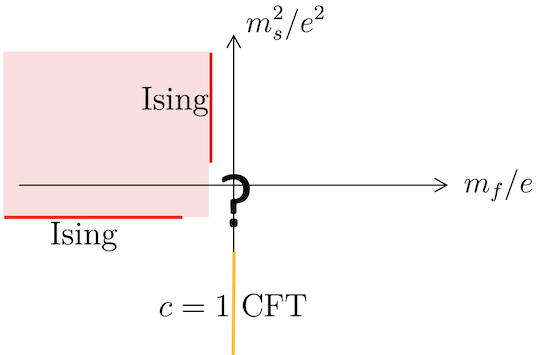}
\end{center}
\caption{The semi-classical limit of the phase diagram.  The shaded red region in the upper-left corner has two ground states; everywhere else has a single ground state. The two lines of Ising $c=1/2$ critical points are shown in red and the $c=1$ critical point in orange.}
\label{phasefig}
\end{figure}

\para
We start by considering the case with $\theta=0$ in the UV. When  $m^2_s\gg e^2$  we can integrate out the scalar. We know from the discussion above that QED$_2$ will have a phase transition as the mass of the fermion moves from positive to negative. This is shown as the vertical red line in Figure \ref{phasefig}.  Meanwhile, if  we integrate out the fermion with $m_f<0$ then we get an effective $\theta=\pi$  in the infra-red and so, as also explained above, we know that there must be a phase transition as $m^2_s$ varies from positive to negative. This is shown as the red horizontal line in Figure \ref{phasefig}. These two lines separate the phase with two ground states, arising from broken charge conjugation, shown as the shaded red region in the top left quadrant, from the phase with a single ground state.  

\noindent

\para
Classically there is a third line of gapless modes. This occurs in the Higgs phase when $m_s^2\ll -e^2$. This gaps both the gauge field and the scalar, leaving behind the fermion. If we additionally set $m_f=0$ then we might expect to get a gapless fermion. But do we?

\para
There is, in fact, a general argument that shows that there  {\it has} to be a line of gapless modes in this vicinity, although not necessarily at $m_f=0$. This follows from the global $U(1)_V$ symmetry \eqn{u1v}. The charges of the fermion and scalar under the gauge symmetry $G$ and global symmetry $V$ are $(G,V) = (1,1)$ for the fermion and $(G,V)=(1,0)$ for the scalar. Consider the theory deep  in the Higgs phase. The global  symmetry $V$ survives and we can introduce an associated background gauge field $\hat{A}$. If we also have $|m_f|\gg e$, then we can integrate out the fermion. When  $m_f>0$ this gives a trivial gapped phase, but when $m_f<0$ this gives an SPT which can be identified by the theta angle for the a background theta term
\be  {\cal L}_{\rm SPT} = \frac{\hat{\theta}}{2\pi} d\hat{A}\ .\ee
In the phase $m_f>0$ we have $\hat{\theta}=0$ and in the phase $m_f<0$, we have $\hat{\theta}=\pi$. Charge conjugation sends $\hat{\theta}\to - \hat{\theta}$, and is unbroken for sufficiently large negative $m_s^2$. Because these two regions lie in different SPT phases, there has to be a phase transition between them where the fermion becomes massless.

\para
The upshot of this semi-classical analysis is shown in  Figure  \ref{phasefig}. In the Higgs phase, $m^2_s\ll -e^2$, we have a line of $c=1$ critical points shown as the orange line.     Elsewhere, we have two lines of $c=1/2$ Ising transitions, shown in red.  Note, in particular that the gapless modes carry the $U(1)$ global charge along the $c=1$ line but along the $c=1/2$ lines all global charge is carried by gapped modes. 

\para
All three lines of fixed points that stretch to the semi-classical region rely on taking $\theta=0$ in the UV. These fixed points are then  repeated at $\theta=\pi$, but reflected in the $y$-axis. When $\theta\neq 0,\pi$,  and so charge conjugation is explicitly broken, there are no phase transitions in this semi-classical region and it seems most plausible that the theory is gapped throughout the $(m^2_s,m_f)$-plane.


\subsection{The Phase Diagram}\label{sfsec}

The next question is: how do the lines in Figure \ref{phasefig} meet up? We can build some intuition by looking more closely at the line of $c=1$ fixed points and following it upwards. 

\para
It will prove useful to work in the bosonised picture \cite{Coleman:1974bu}. This is a standard tool for studying the dynamics of 2d gauge theories but there are various subtleties in bosonisation involving $\Z_2$ gauge symmetries. We will ignore these for now because they don't change our immediate conclusions, but we will deal with bosonisation more carefully  in Section \ref{twofermionsec}.

\para
The standard dictionary bosonises  the fermionic current as  
\be \bar{\psi}\gamma^\mu \psi = -\frac{1}{2\pi} \epsilon^{\mu \nu}\partial_\nu \varphi = -\frac{1}{\pi} \partial^\mu \tvp\,.\ee
Here we have introduced both the scalar $\varphi$ and its dual $\tilde{\varphi}$. (The specific coefficient relating the two holds only at the free fermion point in the $c=1$ moduli space.) Both are compact scalars, with periodicity $\varphi \sim \varphi+2\pi$ and $\tvp\sim \tvp+2\pi$. We will retain this convention of using $\varphi$ and $\tilde{\varphi}$ to denote the bosonic fields associated to the fermion $\psi$ throughout this paper.

\para
We sit in the Higgs phase and  write $\phi = ve^{i\sigma}$. Setting $m_f=0$, the action is
\be 
{\cal L} = \frac{1}{2e^2} F_{01}^2  + v^2 \left(\partial_\mu \sigma +  A_\mu\right)^2  + \frac{1}{8\pi}(\partial_\mu \varphi)^2 +\frac{1}{2\pi} \varphi F_{01}\ .\label{sf}
\ee
With the normalisation chosen  so that $\varphi$ is $2\pi$ periodic, the radius of the compact boson is  determined by the coefficient in front of the kinetic term in  \eqref{sf} and, as expected, matches the free fermion point.  We can now dualise the scalar $\sigma$: we write the dual scalar as $\tilde{\sigma}$ and get the alternative action
 \be {\cal L} = \frac{1}{2e^2} F_{01}^2 + \frac{1}{16\pi^2v^2} (\partial_\mu \tilde{\sigma})^2   + \frac{1}{8\pi} (\partial_\mu {\varphi})^2 + \frac{1}{2\pi}(\tilde{\sigma} + {\varphi}) F_{01}\ .\label{dualscalar}\ee
Here it's clear that there is a gapless mode: the mass comes from integrating out $F_{01}$ but this couples only to the combination $\tilde{\sigma} +{\varphi}$, leaving the other combination $\tilde{\sigma} - {\varphi}$ gapless. Alternatively, if we set  $\tilde{\sigma} = -{\varphi}$ then we see that the gapless mode is given by
\be {\cal L}_{\rm gapless} &=& \frac{1}{8\pi}\left(1+ \frac{1}{2\pi v^2}\right) (\partial_\mu{\varphi})^2\nn\\
&=& \frac{1}{2\pi}\left(1+\frac{1}{2\pi v^2}\right)^{-1} (\partial_\mu \tilde\varphi)^2\ . \label{goingdown}\ee
We also see that the Higgs \vev changes the radius of the scalar, moving us along the $c=1$ conformal manifold\footnote{The fact that $\varphi$  in \eqn{goingdown} remains $2\pi$ periodic is important. It's straightforward to derive this in the present case, but similar issues will arise later where things are a little more subtle. For that reason it is useful to note a more general result. Given a pair of scalars $(\varphi_1,\varphi_2)$ subject to the periodicity conditions $(\varphi_1,\varphi_2)\sim (\varphi_1+2\pi,\varphi_2) \sim (\varphi_1,\varphi_2+2\pi)$, the linear combinations $(\hat{\varphi}_1,\hat{\varphi}_2) = M(\varphi_1,\varphi_2)$ satisfy the same periodicity conditions, $(\hat{\varphi}_1,\hat{\varphi}_2) \sim (\hat{\varphi}_1+2\pi,\hat{\varphi}_2) \sim (\hat{\varphi}_1,\hat{\varphi}_2+2\pi)$, if and only if $M$ is an integer matrix with $\det M=\pm 1$. (Equivalently, $M$ must be an integer matrix with integer inverse.) One can use a field redefinition of this type to verify the result \eqn{goingdown}.}. 

\para
This gives us an important clue. As we move along the $c=1$ line of fixed points, the radius $R$ of the compact boson changes,  which is equivalent to varying the exactly marginal operator of the theory. We work in the convention where the radius is defined as $\calL = \frac{R^2}{4\pi} (\partial {\varphi})^2$ in \eqref{goingdown}. At $v^2\ra \infty$, we have $R^2 = 1/2$, the radius of a free massless fermion. As we decrease $|m_s^2|$, we decrease $v^2$ and hence, from \eqn{goingdown}, increase the radius. 

\para
(Strictly speaking, our initial Lagrangian (\ref{sf}) only describes the Higgs phase of the theory (\ref{1and1}) when suitably coupled to a $\Z_2$ gauge field \cite{arf}. This $\Z_2$ gauging persists in the fixed point theory (\ref{goingdown}), such that we are actually moving along the Dirac branch of the conformal manifold of $c=1$ fermionic CFTs.)

\para
What is the fate of this line of fixed points? Changing the radius of a $c=1$ CFT changes the dimension of the operators of the theory. The stability of this fixed point will change when we hit the radius for which a scalar operator, which is neutral under the $U(1)_V$  flavour symmetry, crosses from being irrelevant to being relevant.


\para
In the language of \eqref{dualscalar}, the $U(1)_V$ symmetry acts as a shift of $\tvp$ and leaves $\varphi$ invariant. This can be seen by including the mass term $m_f \bar \psi \psi$ when bosonising the theory; this generates a $\cos \varphi$ term in \eqref{dualscalar}, which is neutral under  $U(1)_V$. The operator $\bar{\psi}\psi\sim \cos{\varphi}$ is relevant, but it is tuned away by dialing the mass $m_f$. It is the remaining scalar operators that are neutral under $U(1)_V$ that are of concern: these take the form $\cos n \varphi $ for $n\in \Z$ and have dimension $\Delta_n = n^2/(2 R^2)$.

 \begin{figure}[htb]
\begin{center}
\epsfxsize=2.5in\leavevmode\epsfbox{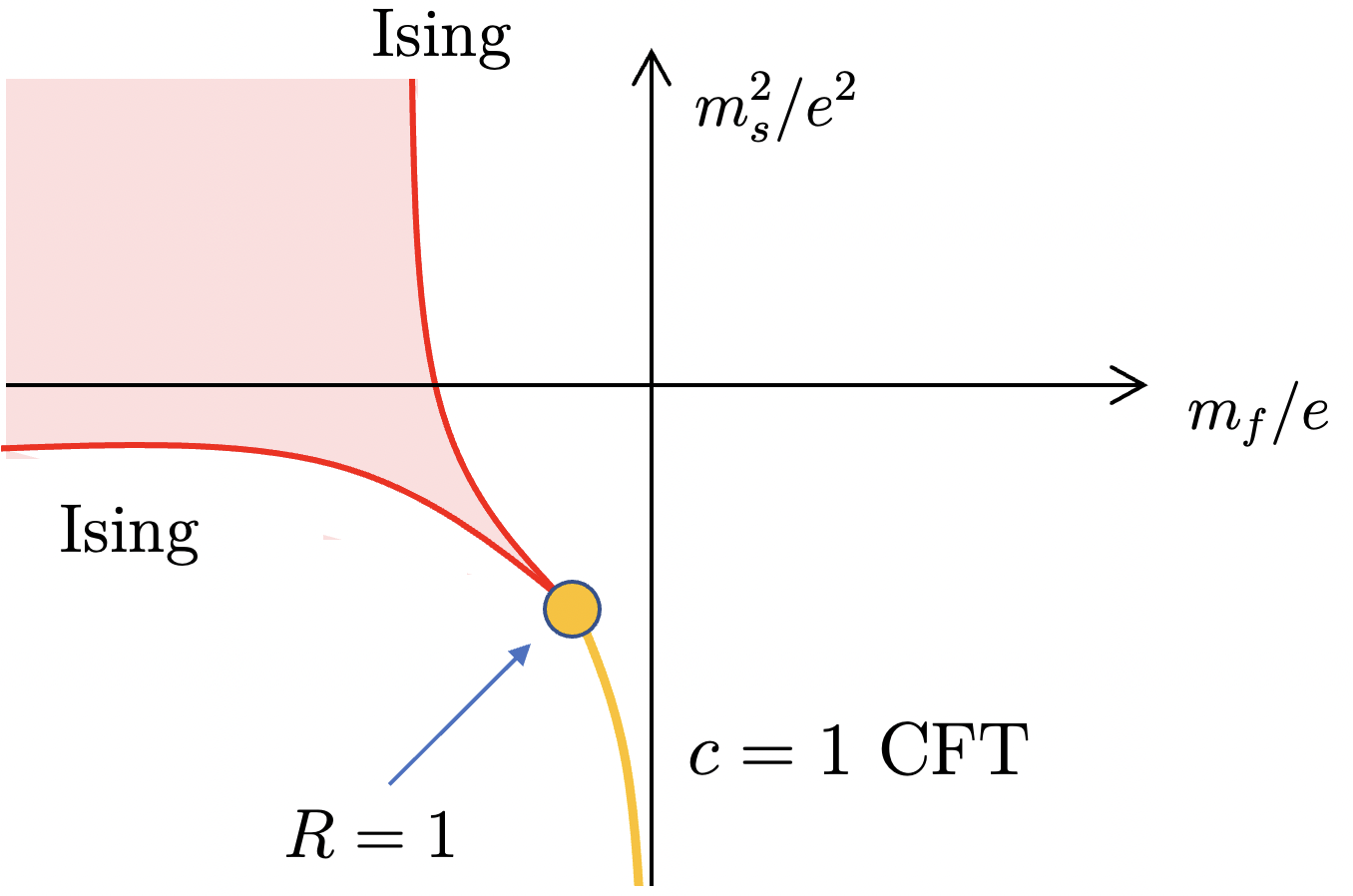}
\end{center}\vspace{-3mm}
\caption{A conjectured completion of the phase diagram: the $c=1$ line ends at the $R=1$ point, where it splits into two $c=1/2$ lines.}
\label{phase2fig}
\end{figure}
\noindent

\para
At the free fermion point, $R^2=1/2$, the only relevant singlet operator is  the mass term $\cos \varphi $. As we move away from the deep Higgs phase, the radius increases, but the stability of the fixed point remains unchanged as long as there are no further relevant operators. Therefore we have a line of $c=1$ CFTs which have a $U(1)_V \times U(1)_A$ symmetry; while the $U(1)_V$ symmetry is present at all energies, the $U(1)_A$ is an emergent symmetry of the IR fixed point only.

\para
The stability of the fixed point changes when we hit the radius $R = 1$, where the operator $\cos 2\varphi $ crosses marginality. \footnote{Note that at $R=1$ on the bosonic $c=1$ moduli space, both $U(1)$ symmetries are enhanced to $SU(2)$. No such symmetry enhancement occurs however for the fermionic theory, since the vertex operators corresponding to the additional currents at $R=1$ are non-local in the fermionic theory.}
Once we have two relevant operators, but only a single parameter $m_f$ to tune,  the $c=1$ fixed point is not stable anymore. We therefore expect the line of $c=1$ fixed points to end at $R=1$. Note, however, that the semi-classical analysis \eqn{goingdown} is valid only for $v^2\gg 1$ and is not sufficient to determine the value of $v^2$ at which $R=1$; this is a problem of strong coupling. 

\para
The SPT argument described previously means that there must be {\it some} phase transition as we pass from the lower-left to the lower-right region of the phase diagram, so it's not possible that the $c=1$ line just ends. We propose that once we reach the $R=1$ point, the line of $c=1$ fixed points splits into two Ising lines, which then connect to the other semiclassical limits of the phase diagram. The result is shown in Figure \ref{phase2fig}.

\subsubsection*{From One $c=1$ to Two $c=1/2$}

Once the $c=1$ line reaches $R=1$, we need to worry about the two operators which might destabilise the fixed point. One is the relevant operator $\calO = \cos \varphi$, with dimension $\Delta = 1/2$, and the other  is a marginally relevant operator which is the linear combination of $\cos 2 \varphi $ and $(\partial \varphi)^2$, with dimension $\Delta = 2$. The IR theory can be obtained by perturbing the $c=1$ self dual point by these two operators
\be
\mathcal{L} = \mathcal{L}_{R = 1} +g \int d^2x\, \calO_{\Delta = 1/2}+\lambda \int d^2 x\,\calO_{\Delta = 2}  \ .\label{eq:double_freq}
\ee
In the scenario we propose, by tuning the values of $g$ and $\lambda$, we could find ourselves in a gapped phase with a single vacuum, a gapped phase with two degenerate vacua, or an Ising CFT. 

\para
The Lagrangian \eqref{eq:double_freq} is called the double frequency Sine-Gordon model \cite{Delfino:1997ya}, and has a $\mathbb{Z}_2$ symmetry $\varphi \to -\varphi$. In our context, this symmetry starts life as the charge conjugation symmetry of the UV Lagrangian. It was shown in \cite{Mussardo:2004rw,Takacs:2005fx} that, depending on the ratio $g/\lambda$, this symmetry could be unbroken or spontaneously broken, with the two phases separated by a second-order phase transition which belongs to the Ising universality class. 

\para
A similar phenomenon happens in the two-flavour Schwinger model which, if both fermions are massless, is described by the compact boson at the self-dual radius \cite{coleman}. Adding small masses implies a perturbation of the kind  \eqref{eq:double_freq}, and it was shown numerically that there is an exponentially small wedge of parameter space where the  $\mathbb{Z}_2$ symmetry is spontaneously broken \cite{bernardo}. This region is separated from the region with an unbroken $\mathbb{Z}_2$ by lines of Ising CFTs. We propose that the same phenomenon occurs for  QED with a scalar and a fermion, as  shown in Figure \ref{phase2fig}.

\section{QED with Two Fermions}\label{twofermionsec}

In this section we turn to the two-flavour Schwinger model: we want to understand QED$_2$, coupled to two Dirac fermions with charges $p$ and $q$, which we take to be co-prime so the gauge symmetry acts faithfully.   If $p$ and $q$ are both odd then the theory is bosonic in the sense that $(-1)^F$ is part of the gauge group. If one is odd and the other even then the theory is fermionic.

\para
We can give a mass to each fermion but, in contrast to the theory discussed in Section \ref{scalarfermionsec}, the point $m_1=m_2=0$ is special because the theory enjoys an enhanced global axial symmetry.

\begin{figure}[htb]
\begin{center}
\epsfxsize=1.8in\leavevmode\epsfbox{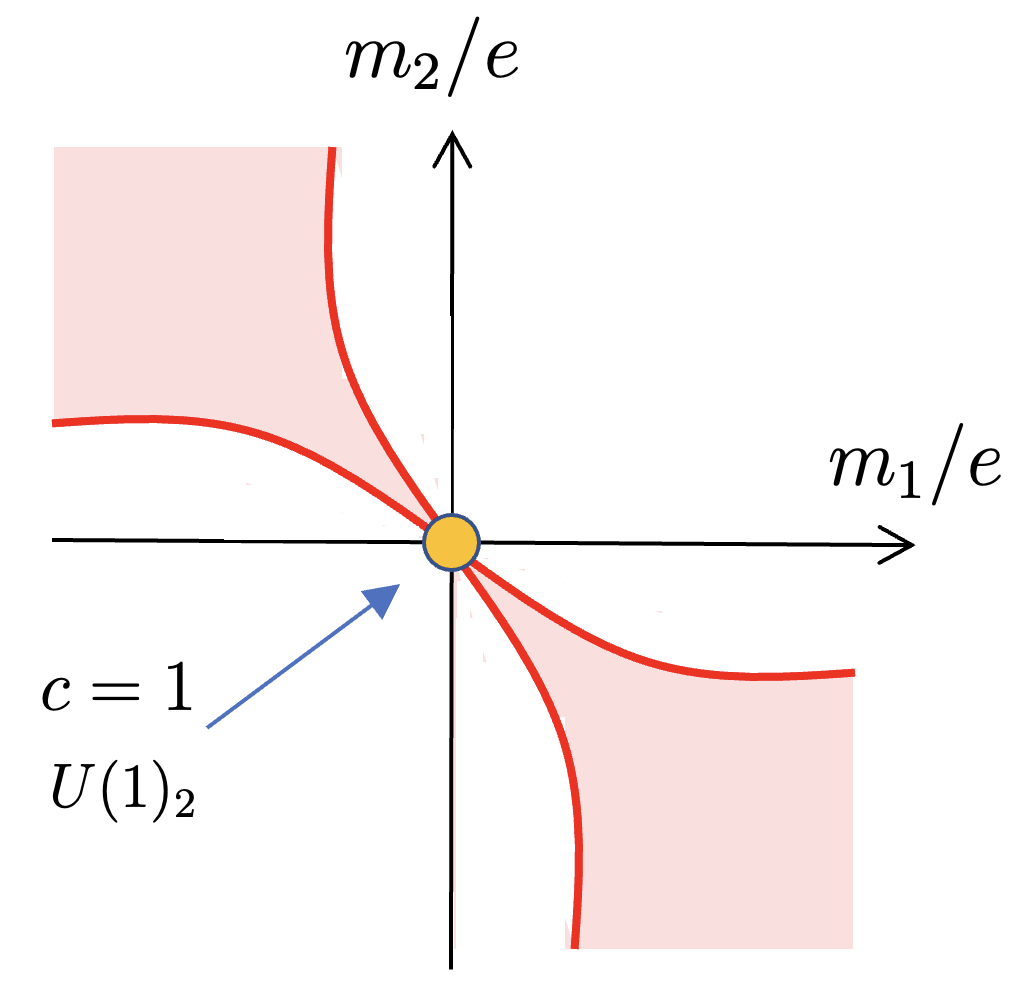}
\end{center}
\caption{The phase diagram for $U(1)$ coupled to two fermions of charge $q=+1$, as derived in \cite{bernardo}. The red shaded region has spontaneously broken charge conjugation and two degenerate ground states.}
\label{phaseqed2fig}
\end{figure}

\para
The case of $p=q=1$ was studied in detail in  \cite{bernardo} and the resulting phase diagram is shown in Figure \ref{phaseqed2fig} (shown for $\theta=0$).  By looking at various limits $m_1,m_2\gg e$ and $m_1,m_2\ll -e$ we can identify lines of Ising phase transitions, separating the region with spontaneously broken charge conjugation (shown in red) from the region with a single ground state. We know that at the origin, where $m_1=m_2=0$, the theory flows to $SU(2)_1 = U(1)_2$ conformal field theory, as shown by the orange dot. 

\noindent

\para
Here our task is to generalise these results to  co-prime charges $p$ and $q$.

\subsection{The Schwinger Model Revisited}\label{twistsec}

Suppose that we give the fermion of charge $p$ a mass $m_1 \gg e$ and integrate it out. We are left with the single-flavour Schwinger model, but with a charge $q$ that, generically, is $q\neq 1$. 

\para
This means that, at low energies, the gauge symmetry does not act faithfully and there will be a $\Z_q$ one-form symmetry, associated  to the $q$ different Wilson lines that cannot end.  There is also a non-anomalous $\Z_q$  0-form global symmetry at $m_2=0$. If we decompose the Dirac fermion into left- and right-moving Weyl fermions, which we write as $\psi_-$ and ${\psi}_+$ respectively, then this $\Z_q$ 0-form symmetry can be taken to act as  $(\psi_-,\psi_+)\to (e^{2\pi i/q}\psi_-,\psi_+)$.

\para
The expectation is that this $\Z_q$ 0-form symmetry is spontaneously broken, and the theory has $q$  distinct ground states. 
As a warm-up, we will see how to reproduce the $q$-fold degeneracy using bosonisation. This is one of the places where we need to do bosonisation properly, replete with the $\Z_2$ gauging, to get to the right answer. Getting things right now will set in good stead for the more subtle chiral theories that we meet later.

\para
To start, we can see where the subtlety arises. The vector and axial symmetries of the fermion map to the winding and shift symmetries of a compact boson. The dictionary is\footnote{In our conventions, the vector symmetry acts as $\psi\to e^{-i\alpha}\psi$ and has current $j_V^\mu = \bar{\psi} \gamma^\mu \psi $, while the axial symmetry acts as $\psi\to e^{-i\alpha \gamma^3}\psi$ and has current $j_A^\mu = \bar{\psi}\gamma^\mu \gamma^3 \psi$. The shift symmetry of the compact boson at radius $R$ acts as $\varphi\to \varphi + \alpha$ and has current $j_\text{shift}^\mu = \frac{R^2}{2\pi} \partial^\mu \varphi$, while the current for the topological winding symmetry is $j_\text{winding}^\mu = \frac{1}{2\pi}(\star d \varphi)^\mu =  \frac{1}{2\pi}\epsilon^{\mu\nu}\partial_\nu \varphi $, which acts as $\tilde{\varphi}\to \tilde{\varphi}+\alpha$ on the dual scalar. The easy way to see (\ref{eq: current matching}) is to act with these symmetries on each side of the bosonisation dictionary $\psi_\pm \leftrightarrow e^{\pm i\varphi/2 + i \tilde{\varphi}}$ valid at $R^2=1/2$. The more careful way is to compute the fermionic currents in terms of bosons using point-splitting regularisation, which yields the same result.}
\begin{align}
  j_A  \,\, \longleftrightarrow \,\,2j_\text{shift}\ \ \ {\rm and}\ \ \ j_V \,\, \longleftrightarrow\,\, -j_\text{winding} \ .
\label{eq: current matching}
\end{align}
That factor of 2 in the first dictionary entry will prove to be important.

\para
For the Schwinger model, there are two different paths to bosonisation. The first, and most obvious, path is to bosonise the charge $q$ Dirac fermion $(\psi_-,\psi_+)$. In this case, we get a bosonised theory written in terms of a compact boson $\varphi\sim\varphi+2\pi$,\footnote{For compactness, here and in what follows we write Lagrangian in terms of forms with $F^2 = \frac{1}{2}F_{\mu\nu}F^{\mu\nu} \,d^2x = -F_{01}^2 \,d^2x$ and $(d\varphi)^2 = \partial_\mu \phi \partial^\mu \phi \,d^2x$.}
\begin{align}
  {\cal L } =  -\frac{1}{2e^2}F^2 +\frac{1}{8\pi}(d \varphi)^2 - \frac{q}{2\pi}\varphi\, dA \ .
\label{1}
\end{align}
The charge $q$ of the fermion appears in the final term. We can now integrate out $A$. One has to take care here to sum over all magnetic flux sectors, the details of which are given in Appendix \ref{app: integrating out gauge fields}. The upshot is that a mass is generated for $\varphi$, with vacua\footnote{One should really think of each of these ground states as the unique ground state of its own \textit{universe}; we'll have more to say about this in the following section.} given by the $q$ constant configurations
\be \varphi = \frac{2\pi j}{q}\ \ \ {\rm with}\  j=0,1,\ldots,q-1\ .\label{v1}\ee
These are the expected $q$ ground states. So far, no surprise. 

\para
Suppose instead that we form the Dirac fermion $(\psi_-,\psi_+^\dagger)$. We now write down a bosonised description directly in terms of the dual scalar $\tilde{\varphi}\sim\tilde{\varphi} + 2\pi$, which reads
\begin{align}
  {\cal L} = -\frac{1}{2e^2}F^2 +\frac{1}{2\pi}(d \tilde{\varphi})^2 + \frac{2q}{2\pi}\tilde{\varphi}\, dA \ . 
\label{2}
\end{align}
Now the problem is clear: that final term has a factor of $2q$ instead of $q$. The additional factor of 2 can be traced to the factor appearing in the current identifications (\ref{eq: current matching}). Integrating out $A$ now imposes the constraint
\be \tilde{\varphi} = \frac{2\pi j}{2q}\ \ \ {\rm with}\ \ j=0,1,\ldots,2q-1\ .\label{v2}\ee
This suggests that the theory has $2q$ vacua. This suggestion is wrong!

\para
What we've missed is the need to include a $\Z_2$ gauging when performing bosonisation. We will discuss this in more detail below, but we can sketch the key idea here. When we work with the original scalar $\varphi$, the $\Z_2$ gauge symmetry is a subgroup of $U(1)_{\rm winding}$. As such, it doesn't affect the constant ground states \eqn{v1}. In contrast, when we work with the dual scalar $\tilde{\varphi}$, the $\Z_2$ gauge symmetry is a subgroup of $U(1)_{\rm shift}$. That means that the gauge symmetry relates the  vacua
\begin{align}
    j  \,\, \longleftrightarrow \,\, j+q \ ,
\end{align}
for each $j=0,\dots,q-1$. To see how this works in more detail, we to turn to the partition function of the theory.

\subsubsection*{The View From the Partition Function}

It's not true that a fermion in 2d is the same as a boson. Instead, a fermion is the same a boson coupled to a $\Z_2$ gauge field, and vice versa. This is most simply seen by examining the partition functions of the two theories.

\para
We start with a compact boson $\varphi\sim \varphi+2\pi$ of radius $R$ which, in our conventions, means we have action
\begin{align}
  S = \int \frac{R^2}{4\pi}(d \varphi)^2 \ .
  \label{eq: generic boson action}
\end{align}
T-duality acts as $R\leftrightarrow 1/R$, and exchanges the two global symmetries $U(1)_\text{shift}\leftrightarrow U(1)_\text{winding}$.

\para 
At $R^2=1/2$, this theory describes a single complex Dirac fermion. However, the local operators of one theory correspond generically to twist operators of the other, which are not local but instead live at the end of some $\Z_2$ topological line. In terms of the Hilbert space on ${\bf S}^1$, states in the untwisted sector of one theory may be in a $\Z_2$ twisted sector of the other. 

\para
We take the compact boson of radius $R$ on a Euclidean torus with modular parameter $\tau=\omega_2/\omega_1$. Next, we  introduce a $\Z_2$ twist\footnote{These twists are enacted in the path integral by the insertion of the network of topological defects
\begin{align}
  \D_{\alpha,\beta} = \exp \left(\frac{i \alpha}{2}\int_{\omega_2}d\varphi\right)\, \exp \left( \frac{i \beta}{2}\int_{\omega_1}d\varphi\right)\ .
\end{align}
} by $(-1)^\alpha\in  U(1)_\text{winding}$ around the $\omega_1$ cycle and by $(-1)^\beta\in  U(1)_\text{winding}$ around the $\omega_2$ cycle, where $\alpha,\beta\in \{0,1\}$. We then write the corresponding twisted partition function as $Z_\text{w}^{(\alpha,\beta)}(R)$, given by
\begin{align}
  Z_\text{w}^{(\alpha,\beta)}(R) = \frac{1}{\eta \bar{\eta}}\sum_{k,m\in \Z} (-1)^{\beta m} \zeta^{(k+mR^2+\alpha /2)^2/4R^2}\bar{\zeta}^{(k-mR^2+\alpha /2)^2/4R^2} \ ,
  \label{eq: twisted pfs}
\end{align}
where $\zeta=e^{2\pi i \tau}$ and $\eta=\eta(\tau)$ is the Dedekind $\eta$-function. (The variable $\zeta$ is more usually called $q$, but that name is already taken as the charge of our fermion!)

\para
Alternatively, we could consider a $\Z_2$ twist but now in the $U(1)_{\rm shift}$ symmetry instead of the winding symmetry.  This is implemented by imposing the twisted boundary conditions,
\begin{align}
  e^{i\varphi(z+\omega_1)} = (-1)^\alpha e^{i\varphi(z)}\ \ \ {\rm and}\ \ \  e^{i\varphi(z+\omega_2)} = (-1)^\beta e^{i\varphi(z)} \ .
\end{align}
We denote the corresponding partition function by $Z_\text{s}^{(\alpha,\beta)}(R)$.  T-duality ensures the two partition functions are related by 
\begin{align}
  Z_\text{w}^{(\alpha,\beta)}(R) = Z_\text{s}^{(\alpha,\beta)}(1/R) \ .
  \label{eq: twisted T-duality}
\end{align}
A complex Dirac fermion is usually said to be equivalent to a compact scalar at radius $R^2=1/2$. However, the correct statement is that a complex Dirac fermion is equal to a sum over the bosonic partition functions. This is what is meant by gauging the $\Z_2$ symmetry: we sum over all holonomies or, equivalently, all twisted sectors. In detail, at generic radius $R$ we can define
\begin{align}
  Z_\text{Dirac}(R) 	&= \frac{1}{2}\sum_{\alpha,\beta}(-1)^{\alpha\beta} Z_\text{w}^{(\alpha,\beta)}(R) 		\nn\\
  						&= \frac{1}{\eta\bar{\eta}} \sum_{n,m\in \Z} \zeta^{((n+m)/R+2(n-m)R)^2/16} \bar{\zeta}^{((n+m)/R-2(n-m)R)^2/16} \ .
\label{eq: Dirac branch}
\end{align}
As we vary $R$, this defines the NS-NS partition function of every theory on the `Dirac' branch\footnote{This is to be contrasted with the orbifold branch, obtained by taking an orbifold by $\varphi \to - \varphi$. In the fermionic language, this corresponds to orbifolding by charge conjugation.} of the moduli space of $c=1$ fermionic CFTs \cite{arf}. The sign $(-1)^{\alpha\beta}$ can be understood in the language of Arf invariants. From the perspective of the Hilbert space on the $\omega_1$ cycle, say, it tells us to project onto states that are even under $\Z_2\subset U(1)_\text{winding}$ in the untwisted sector, but \textit{odd} under this symmetry in the twisted sector. One can equivalently use the T-duality relation (\ref{eq: twisted T-duality}) to understand $Z_\text{Dirac}(R)$ as a gauging by $\Z_2\subset U(1)_\text{shift}$.

\para
Note that we have the duality
\begin{align}
  Z_\text{Dirac} (R) = Z_\text{Dirac}(1/2R)\ ,
\label{eq: fermionic T-duality}
\end{align}
which should be contrasted with the T-duality of the bosonic theory under $R\leftrightarrow 1/R$. At the self-dual point $R^2=1/2$ we find the NS-NS partition function of the free Dirac fermion,
\begin{align}
  Z_\text{Dirac}(1/\sqrt{2})=\frac{1}{\eta \bar{\eta}} \sum_{n,m\in \Z} \zeta^{n^2/2}\bar{\zeta}^{m^2/2}\ .
\end{align}
We will later find theories which at low energies land at different points of this line of $c=1$ fermionic CFTs.
\para 
Now we can see how these partition functions play out in our two choices for bosonisation. First, if we take the standard route to bosonisation, so that we are left with the Lagrangian \eqn{1}, then the fermionic partition takes the form
\be Z = \frac{1}{2}\sum_{\alpha,\beta}(-1)^{\alpha\beta} Z_\text{w}^{(\alpha,\beta)}(1/\sqrt{2})\ .\ee
As we've seen, when we integrate out the gauge field, the bosonic theory has $q$ vacua. This means that, at low-energies, $Z_\text{w}^{(\alpha,\beta)}\ra q$ and so the fermionic partition function becomes
\be Z \ra \frac{1}{2}\left(q+q+q-q\right) = q \ .\ee
These are the  expected $q$ ground states in a theory with a $\Z_q$ one-form symmetry. 

\para
We can also see what happens if we take the other route \eqn{2} to bosonisation. Now there are naively $2q$ ground states, which we know is the wrong answer. But, because we're working the dual scalar $\tilde{\varphi}$, the $\Z_2$ gauge symmetry lies in the shift symmetry $\Z_2 \subset U(1)_{\rm shift}$. The fermionic partition function is
\be Z = \frac{1}{2}\sum_{\alpha,\beta}(-1)^{\alpha\beta} Z_\text{s}^{(\alpha,\beta)}(\sqrt{2})\ .\ee
However, the constant vacua \eqn{v2} that we found for the compact boson are only valid configurations in the untwisted sector.  This means that, at low energies, we have
\begin{align}
  Z_\text{s}^{(\alpha,\beta)} \to  \left\{\begin{aligned}
  	&2q		\qquad 	&&\text{if }\alpha = \beta = 0		\\
  	&0				&&\text{otherwise}
  \end{aligned}\right. \ ,
\end{align}
and hence
\begin{align}
  Z\to \frac{1}{2}\left(2q+0+0-0\right) = q \ ,
\end{align}
which is the correct answer.

\para
For the single flavour Schwinger model, we had two ways to do bosonisation and, by picking the choice \eqn{1}, we can brush the issue of the $\Z_2$ gauge symmetry under the carpet. We will not have the same luxury when we turn to chiral theories in Section \ref{chiralsec} and the kind of analysis that we've done above will be needed.

\subsection{Two Flavour Schwinger Model}\label{multisec}

Now we can turn  to our real interest in this section:  $U(1)$ gauge theory coupled to two Dirac fermions, with co-prime charges $p$ and $q$. This is a non-chiral gauge theory.

\para
The (invertible, 0-form) global symmetries of the theory make up $U(1)^2$.  We have two left-movers and two right-movers, with charges\footnote{
Note that some elements of $\widehat{G}_1\times \widehat{G}_2$ act trivially up to a gauge transformation. The faithfully acting symmetry is $G\times \widehat{G}_1\times \widehat{G}_2/(\Z_k\times \Z_2)$, with $\Z_k$ generated by $(e^{2\pi i p/k},e^{2\pi i q/k},1)$ and $\Z_2$ generated by $(1,-1,-1)$
}
\begin{equation}
 	\begin{array}{c|cccc}
&\psi_{-1}&\psi_{-2} & \psi_{+1} & \psi_{+2} \\ \hline
G & \,p\ & q & \ p  & q\\
\widehat{G}_1	& q\ & -p & \  q & -p		\\
\widehat{G}_2	& q\ & -p & \  -q & p		\\
\end{array}\
\label{pqinnit} \end{equation}
 under the gauge symmetry $G$ and global symmetries $\widehat{G}_1,\widehat{G}_2$. We additionally have a $\Z_2$ charge conjugation symmetry. In the special case $p=q=1$, both global symmetries are enhanced to $SU(2)$.

\para
In the absence of any mass terms, the theory flows to a $c=1$ CFT. (The story for a general non-Abelian gauge group was discussed in  \cite{diego1,diego2} where it was conjectured that the theory is given in terms of a particular coset.) That means that the  IR theory can be described in terms of a compact boson (\ref{eq: generic boson action}) for some radius $R$, at least up to subtleties related to the $\Z_2$ gauging.

\para
The obvious question is: where in the moduli space of $c=1$ CFTs do we land  when the charges are $p$ and $q$? Or, in other words, what's $R$?

\para
This is easily answered using bosonisation. We have two compact bosons, $\varphi_1$ and $\varphi_2$, with Lagrangian 
\begin{align}
{\cal L} &= -\frac{1}{2e^2}F^2+\frac{1}{8\pi}(d \varphi_1+2q\hat{A}_2)^2+\frac{1}{8\pi}(d \varphi_2-2p\hat{A}_2)^2  \nn\\
&\qquad - \frac{1}{2\pi}\left(p\varphi_1 + q\varphi_2\right) dA - \frac{1}{2\pi}\left(q\varphi_1 - p\varphi_2\right) d\hat{A}_1 \ ,
  \label{eq: pq vector action}
\end{align}
where we've included background gauge fields $\hat{A}_1,\hat{A}_2$ for $\widehat{G}_1,\widehat{G}_2$, respectively, in addition to the dynamical gauge field $A$. The lack of invariance of the corresponding partition function under gauge transformations of $\hat{A}_2$ when $\hat{A}_1\neq 0$ is a symptom of the mixed `t Hooft anomaly between $\widehat{G}_1$ and $\widehat{G}_2$.

\para 
Integrating out the gauge field $A$ gaps the combination $(p\varphi_1 + q\varphi_2)$. More carefully, we can define a new pair of scalars $(p\varphi_1 + q\varphi_2)$ and $(a\varphi_1 + b\varphi_2)$, which are each independently $2\pi$-periodic if we take $aq-bp= 1$. This is indeed solvable for $(a,b)$ precisely because $(p,q)$ are co-prime. The result is that we're left with a single gapless boson, $\varphi=a\varphi_1+b\varphi_2$, with Lagrangian
\begin{align}
  {\cal L} = \frac{p^2+q^2}{8\pi}(d \varphi + 2\hat{A}_2)^2 - \frac{p^2 + q^2}{2\pi} \varphi \,d\hat{A}_1 \ , \label{eq:Lpq_rad}
\end{align}
with $\varphi\sim \varphi+2\pi$. This gives us the answer that we wanted: the $U(1)$ gauge theory with two Dirac fermions with charges $p$ and $q$ flows to a compact boson with radius 
\be R^2 = \frac{{p^2 +q^2}}{2} \equiv \frac{k}{2}\ .\label{pqradius}\ee
As a sanity check, note that if we take $p=q=1$ then the original gauge theory has an $SU(2)$ global symmetry and, indeed, the radius $R^2=1$  is the self-dual point where the compact boson has an enhanced $SU(2)$ chiral algebra.

\subsubsection*{Fermionic Versus Bosonic Theories}

The answer \eqn{pqradius} for the radius is correct, but we have not yet determined the corresponding partition function. The partition function of the initial theory such that it properly describes the fermions is
\begin{align}
  Z_{p,q} = \frac{1}{4}\sum_{\alpha,\beta,\gamma,\delta}(-1)^{\alpha \beta +\gamma \delta}Z^{(\alpha,\beta;\gamma,\delta)} \ ,
\end{align}
where $Z^{(\alpha,\beta;\gamma,\delta)}$ is the partition function of the action (\ref{eq: pq vector action}) with an $(\alpha,\beta)$ twist in the $U(1)_\text{winding}$ of $\varphi_1$ and a $(\gamma,\delta)$ twist in the $U(1)_\text{winding}$ of $\varphi_2$. Note that the two scalars are coupled through the gauge field, and thus $Z^{(\alpha,\beta;\gamma,\delta)}$ does not factorise.

\para
The basic task at hand is then to keep track of these twists whenever we make field redefinitions, or dualise scalars, or integrate out gauge fields. Let us simply state the result in this case. We find that at low energies we flow to the partition function
\begin{align}
  Z_{p,q} =\frac{1}{4}\sum_{\alpha,\beta,\gamma,\delta}(-1)^{\alpha \beta +\gamma \delta}Z_\text{w}^{(q\alpha - p \gamma,q\beta - p \delta)} \left(\sqrt{k/2}\right)  \ ,
\end{align}
where the twisted partition function $Z_\text{w}^{(\alpha,\beta)}(R)$ was determined in (\ref{eq: twisted pfs}). 

\para
There are then two cases to consider. If $p,q$ are both odd, so that the theory is bosonic, we have
\begin{align}
  Z_{p,q} =\frac{1}{4}\sum_{\alpha,\beta,\gamma,\delta}(-1)^{\alpha \beta +\gamma \delta}Z_\text{w}^{(\alpha -  \gamma,\beta -  \delta)} \left(\sqrt{k/2}\right)= Z\!\left(\sqrt{k/2}\right) \ .
\end{align}
Thus we flow simply to a compact boson of radius $R^2=k/2$, which has a manifestly bosonic spectrum.

\para
If on the other hand one of $p,q$ is even, so the  theory is fermionic, we have
\begin{align}
  Z_{p,q} &=\frac{1}{4}\sum_{\alpha,\beta,\gamma,\delta}(-1)^{\alpha \beta +\gamma \delta}Z_\text{w}^{(\alpha ,\beta )} \!\left(\sqrt{k/2}\right) 		\nn\\
  &=\frac{1}{2}\sum_{\alpha,\beta}(-1)^{\alpha\beta}Z_\text{w}^{(\alpha ,\beta )} \!\left(\sqrt{k/2}\right)	
= Z_\text{Dirac}\!\left(\sqrt{k/2}\right) \ ,
\end{align}
where we defined $Z_\text{Dirac}(R)$ in (\ref{eq: Dirac branch}). Thus, as expected, we land at a particular point in the moduli space of $c=1$ fermionic CFTs.

\subsection{The Phase Diagram}

We now turn to the phase diagram of the two-flavour Schwinger model with charges $(p,q)$ as specified in \eqn{pqinnit}. We are interested in the phase structure as we turn on two masses
\be
\label{2Schwingermassterms}
{\cal L}_{\rm mass} = m_1\left(\psi_{-1}^\dagger\psi_{+1} + \text{h.c.}\right) + m_2\left(\psi_{-2}^\dagger\psi_{+2}+ \text{h.c.}\right)\ .\ee
We can use a non-anomalous axial rotation to show that the phase diagram is symmetric under $(m_1,m_2)\to ((-1)^pm_1,(-1)^q m_2)$.

\para  
Consider first the asymptotic regions of the phase diagram. We can take $m_1\gg e$ and integrate out the Dirac fermion $(\psi_{-1},\psi_{+1})$. We are left with the charge $q$ Schwinger model at $\theta=0$. Ignoring the massive fermion, the theory has a $\Z_q$ 1-form symmetry which tells us that the theory's Hilbert space splits into $q$ \textit{universes}, superselection sectors that are separated by infinite potential barriers. This $\Z_q$ 1-form symmetry is broken by the massive charge $p$ fermion and these infinite barriers become finite, allowing the different universes to communicate with each other. Nonetheless, understanding the physics of these decoupled universes  in the charge $q$ Schwinger model will prove fruitful. The fate of these ground states as we turn on $m_2$ was described in \cite{zohart,misumi,zoharschwinger}.

\para
At $m_2=0$ the theory has a $\Z_q$ 0-form axial global symmetry which is spontaneously broken; we get one degenerate ground state from each of the $q$ universes. For $m_2\neq 0$, this $\Z_q$ 0-form symmetry is explicitly broken and the ground state energies of the $q$ different universes are no longer degenerate. What happens depends on whether $q$ is even or odd.

\para 
We described the  $q=1$ theory in Section \ref{qedsec}. There is no ground state degeneracy at $m_2=0$ and there continues to be a unique ground state for all $m_2>0$. In contrast, as we turn on $m_2<0$, we hit a second order phase transition of the  Ising universality class at $m_2\approx - e/3$, after which we have two ground states, reflecting the spontaneous breaking of charge conjugation.

\begin{center}
\centering
\includegraphics[width=110mm]{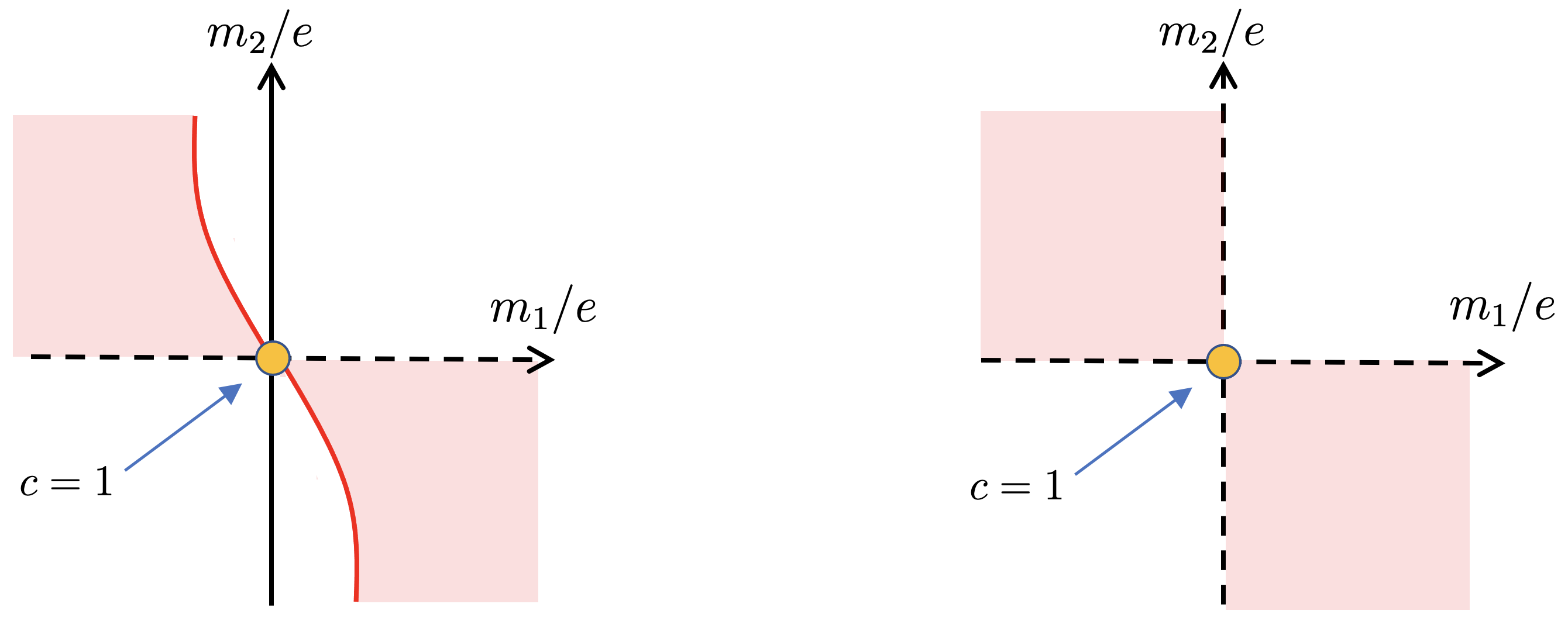}
\captionof{figure}{The phase diagram for $U(1)$ coupled to two fermions of co-prime charges $p$ and $q$ with $pq$ odd. The region of broken charge conjugation  is shaded in red. The red line denotes an Ising transition and a dotted line a first order phase transition. The diagram on the left has $p=1$ and $q\neq 1$ and the diagram on the right has $p,q\neq 1$.}\label{phaseqed22fig}
\end{center}
\noindent

\para 
Next consider $q$ even. In this case, there is a unique lowest energy ground state for $m_2\neq 0$, but which ground state depends on the sign of $m_2$. In more detail, at $m_2=0$ there are two ground states that are invariant under charge conjugation and one of these becomes the true ground state when $m_2>0$ and the other is the true ground state when $m_2<0$. In this way, we have a first order phase transition about $m_2=0$.  We do not expect any further transition, and indeed we find a unique ground state at $|m_2|\gg e$. (The Ising transition that occurs in the $q=1$ theory now  happens in a universe with higher energy that is not the true ground state.)

\para
Things are different for $q>1$ and odd.  There is only one charge conjugation singlet among the ground states at $m_2=0$. For $m_2>0$ this is the unique ground state, while for $m_2<0$ there are two degenerate ground states which spontaneously break charge conjugation. Again, no further transition is expected, matching the single ground state at $m_2\gg e$ and two ground states at $m_2\ll -e$.


\para
The opposite asymptotic regime $m_1\ll -e$ can be dealt with in a similar way. If the charge  $p$  of the massive fermion is even, then we once again find the charge $q$ Schwinger model with $\theta=0$ and the analysis goes through identically. If $p$ is odd we instead have the charge $q$ Schwinger model with $\theta=\pi$. Most of the physics is fixed by the $(m_1,m_2)\to ((-1)^pm_1,(-1)^q m_2)$ symmetry, but there is a small novelty when $p$ is odd and $q$ is even. Now charge conjugation is broken for any $m_2\neq 0$, with two degenerate ground states for both $m_1>0$ and $m_2<0$. For $q\geq 4$, the pair of ground states is different and there is a first order phase transition across the $m_2=0$ axis, but for $q=2$ there are only two ground states at $m_2=0$ which persist for $m_2\neq 0$. This means that there is no first order phase transition for $q=2$.

\begin{center}
\centering
\includegraphics[width=110mm]{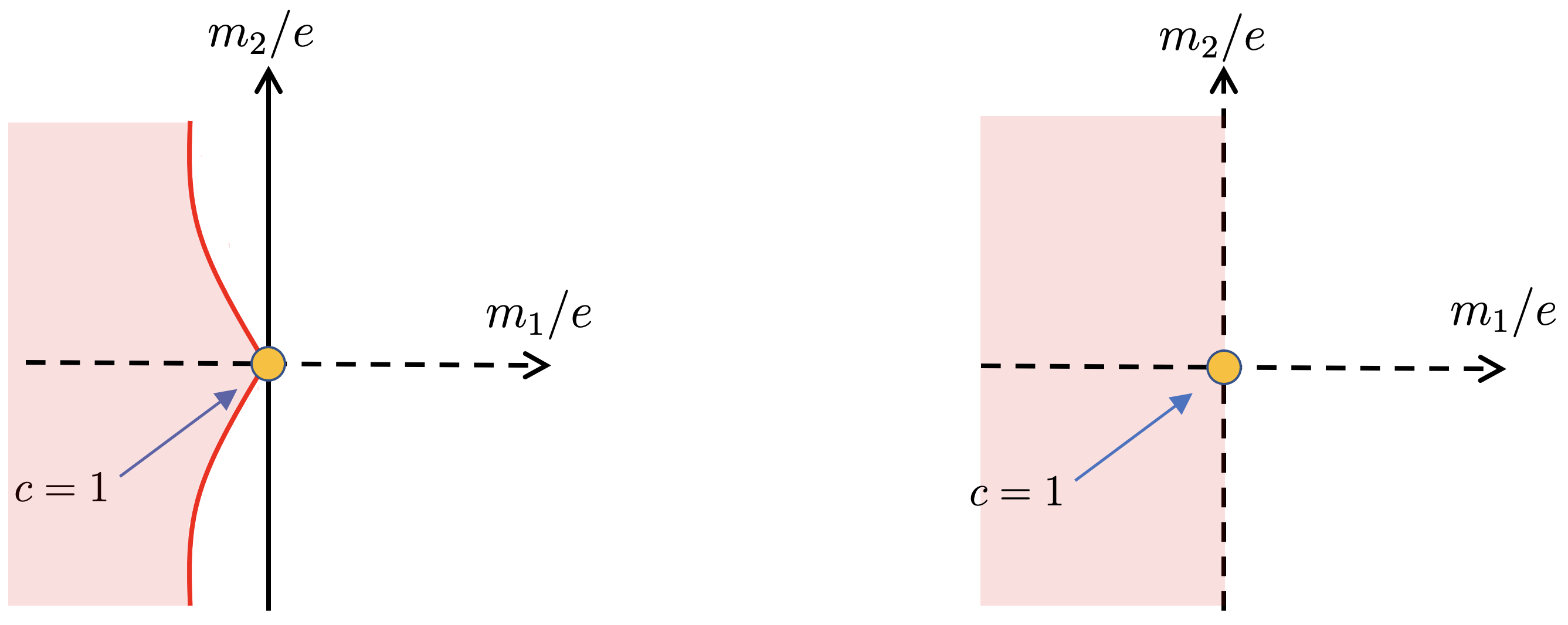}
\captionof{figure}{The phase diagram for $U(1)$ coupled to two fermions of co-prime charges $p$ and $q$ with $q$ even. The region with broken charge conjugation  is shaded in red. The red line denotes an Ising transition and the dotted line a first order transition. The diagram on the left has $p=1$ and the diagram on the right has $p\neq 1$. (The first order transition across the negative $m_1$ axis is absent for $q=2$.)}\label{phaseqed33fig}
\end{center}

\para 
Although the analysis above was done for the Schwinger model with charge $q$, corresponding to the $m_1\gg e$ limit of our theory, the key features persists for all $m_1$. In particular, the $\Z_q$ 0-form symmetry at $m_2=0$ exists for all  $m_1$. To see this, note that while generic $(m_1,m_2)$ completely breaks $\widehat{G}_2$ (and preserves $\widehat{G}_1$), there are distinguished lines of enhanced symmetry: along $m_2=0$ $\widehat{G}_2$ is broken to $\Z_{2q}$, while along $m_1=0$ it is broken to $\Z_{2p}$. The  $\Z_q$ 0-form symmetry of the single-flavour Schwinger model survives for all $m_1$ and is generated
 by $(e^{\pi i/q},e^{\pi i /q})\in \widehat{G}_1\times \widehat{G}_2$. Furthermore this $\Z_q$ symmetry has a mixed anomaly with $\widehat{G}_1$, with anomaly coefficient
\begin{align}
  \mathcal{A} 	&= p^2 + q^2 \,\, \left(\text{mod} \,\,q \right) \ .	\label{anompq}
\end{align}
Since $(p^2 + q^2)$ is co-prime to $q$, any gapped phase must completely break $\Z_q \to 1$. The upshot is that the $q$-fold degeneracy of vacua we found at $m_2=0$ and $m_1\to \infty$ is robust as we reduce $m_1$ and move into the interior of the phase diagram.

\para
The discussion above is already sufficient to sketch the phase diagrams. The phase diagram is shown in Figure \ref{phaseqed22fig} for  $pq$ odd and in Figure \ref{phaseqed33fig} and for $pq$ even. (In the latter case, there is no first order phase transition across the $m_1<0$ axis when $q=2$.)

\para
We can also explore the phase diagram starting from the $c=1$ fixed point at $m_1=m_2=0$. We will see that this confirms the pictures shown in Figures \ref{phaseqed22fig} and \ref{phaseqed33fig}. 
Recall that the fixed point at the origin is described by a $c=1$ bosonic CFT of radius $R^2 = k/2$ when $pq$ is odd, and a $c=1$ fermionic CFT of radius $R^2 = k/2$ when $pq$ is even. (Here $k=p^2+q^2$.) The vertex operators of this fixed point theory are
\be {\cal O}_{m,n} = 
  e^{in\varphi}e^{im\tilde{\varphi}}\ .
  \label{eq: generic vertex operator}
\ee
We have $n,m\in \Z$ when $pq$ is odd, while we have $m\in \Z$ and $n+(m/2)\in \Z$ when $pq$ is even. These have scaling dimensions, spins, and global charges given by 
\begin{align}
  (\Delta,s) = \left(\frac{n^2}{k}+\frac{k m^2}{4},|nm|\right),\ \text{with charges }(km,-2n) \text{ under }(\widehat{G}_1,\widehat{G}_2) \ .
\label{eq: vertex operator quantum numbers}
\end{align}
We see that the $\widehat{G}_1\times \widehat{G}_2$ symmetry ensures the stability of the fixed point: all vertex operators are charged under either $\widehat{G}_1$ or $\widehat{G}_2$.

\para
Suppose that we now turn on $m_1\neq 0$ while keeping $m_2=0$. We retain a $\Z_{2q}\subset \widehat{G}_2$ symmetry along with all of $\widehat{G}_1$. We should expect then any scalar operator that is invariant under both charge conjugation and $\widehat{G}_1\times \Z_{2q}$ to be generated in the Lagrangian; these are $\cos(aq\varphi)$ for $a\in \Z$. For any non-zero $q$, at least the first of these, $\cos(q\varphi)$, is relevant. But, regardless of which combination of these operators is generated, they can only gap the theory\footnote{This is straightforward to see for the bosonic theories. For the fermionic theories, one needs to remember to do the sum over twisted sectors. But indeed we see that each of the four sectors realises $q$ vacua, and so $Z\to \frac{1}{2}(q+q+q-q) = q$.} down to a $q$-fold degeneracy of vacua as they are all invariant under $\varphi\to \varphi +2\pi/q$,  agreement with the anomaly cancellation \eqn{anompq}.

\para
Away from the $m_1=0$ and $m_2=0$ axes, things are less constrained. The $\widehat{G}_2$ symmetry is completely broken and so one can generate $\cos(n\varphi)$ for any $n\in \Z$. We can nonetheless assess the situation very close to the origin by constructing the potential associated to  mass terms (\ref{2Schwingermassterms}). This is 
\begin{align}
  V(\varphi) \sim  -\Lambda\,\big( m_1 \cos(q\varphi) + m_2 \cos(p\varphi)\big) \ ,
\label{eq:mass_def}
\end{align}
where $\Lambda$ is a positive energy scale coming from bosonisation, and we do not write a positive normalisation. The normal ordered operators  $:\cos(q\varphi):$ and $:\cos(p\varphi):$  have dimensions $\Delta = {p^2}/{k}$ and $\Delta = {q^2}/{k}$, respectively, and thus are both always relevant. We can get a qualitative picture of the phase diagram by studying the minima of the potential as we vary $m_1/m_2$.

\para
The case $p=q=1$ is special. In this case, both mass deformations become the operator $\cos \varphi$ with dimension $\Delta= 1/2$, but the masses can be tuned to $m_1 = -m _2$ so that the this operator is tuned away. Then the higher contributions neglected in \eqref{eq:mass_def} become important. It was shown in \cite{bernardo} that for $m_1 = -m_2$ the theory is perturbed by a marginally relevant operator which is a linear combination of $\cos(2\varphi)$ and $(\partial\varphi)^2$. Indeed, it was shown that by tuning the ratio of fermion masses there is an (exponentially small) wedge of phase space in which only $\cos(2\varphi)$ is generated and the theory does indeed have two vacua. The resulting phase diagram was shown in Figure \ref{phaseqed2fig}.

\para
In all other cases, we can take the potential \eqn{eq:mass_def} at face value and explore its ground states as we vary $m_1$ and $m_2$. The resulting phase diagram depends on whether $pq$ is even or odd and agrees exactly with Figure \ref{phaseqed22fig} and Figure \ref{phaseqed33fig}. In this way, the potential \eqn{eq:mass_def} captures both the first and second order phase transitions. (Indeed, in the case $p,q\neq 1$ the potential also captures the Ising transition that is seen in higher energy meta-stable vacua and are not shown in the figures.) We can also use this classical analysis to determine the opening angle between  the Ising line and the $m_2$ axis with $p=1$: it is $\beta=\cot^{-1}(q^2)$.

\para
There is one final consistency check we can do on our proposed phase diagrams. The global symmetry $\widehat{G}_1$ is a good symmetry everywhere in the phase diagram, and so we can turn on a background gauge field for it and determine the corresponding IR background theta angle $\hat{\theta}$ in various regions. Charge conjugation sends $\hat{\theta}\to - \hat{\theta}$, and thus $\hat{\theta}$ cannot change continuously in regions of the phase diagram in which charge conjugation is unbroken. It is then straightforward to compute $\hat{\theta}$ in the asymptotic regions $|m_1|,|m_2|\gg e$. For the bosonic theories shown in Figure \ref{phaseqed22fig}, we have $\hat{\theta}=0$ in the top right and bottom left corners, and $\hat{\theta}=\pi$ in the top left and bottom right corners. It follows that we have $\hat{\theta}=0$ in all unshaded parts of the phase diagrams, while in the shaded regions $\hat{\theta}$ can vary at will, so long as it approaches $\hat{\theta}\to \pi$ as $|m_{1,2}|\to \infty$, and takes the value $\hat{\theta}=0$ along the Ising line of the left diagram.  For the  fermionic theories shown in Figure \ref{phaseqed33fig}, we find asymptotically that $\hat{\theta}=0$ in the top left and top right corners, and $\hat{\theta}=\pi$ in the bottom left and bottom right corners. In both diagrams, the unshaded portion above the $m_1$ axis must have $\hat{\theta}=0$ while the unshaded portion below the $m_1$ axis has $\hat{\theta}=\pi$, with a discontinuous jump at the lines of first order transitions along the positive $m_1$ axis.
\section{Chiral Theories}\label{chiralsec}

In this section, we turn our attention to chiral gauge theories in which left- and right-moving fermions carry different charges.  We will again use bosonisation to understand the dynamics of these theories.

\para
One might think that the way forward is to work with chiral bosons, but these are notoriously subtle objects.  (See \cite{ashoke,frodoandsam} for recent discussions.) It is, however, quite possible to work with the usual non-chiral bosonisation of Dirac fermions, and then gauge a chiral symmetry. This is one situation where the $\Z_2$ gauging inherent in bosonisation becomes important. 

\para
We will ultimately work towards a better understanding of symmetric mass generation, which entails a theory that includes both chiral fermions and Higgs fields. We will describe this in Section \ref{smgsec}. But we start by omitting the Higgs fields and discussing the dynamics of some simple chiral (and, in one case, non-chiral) gauge theories. These will be of some interest in their own right, but also serve as a testing ground to apply bosonisation to compute the number of ground states of a theory, taking into account the $\Z_2$ twists.

\subsection{Chiral QED}

The simplest chiral gauge theory is the 3450 model, a $U(1)$ gauge symmetry $G$ coupled to two left-moving fermions $\psi_{-1}$ and $\psi_{-2}$ with charges 3 and 4, and two right-moving fermions $\psi_{+1}$ and $\psi_{+2}$ with charges 5 and 0. 

\para
This theory also enjoys a non-anomalous global symmetry $\widehat{G}$, with charges
\begin{equation}
 	\begin{array}{c|cccc}
&\psi_{-1}&\psi_{-2} & \psi_{+1} & \psi_{+2} \\ \hline
G & 3 & 4 &  5 & 0\\
\widehat{G}	& 2 & 1 & 2 & 1
\end{array}\label{3450}
 \end{equation}
The dynamics of this theory is largely fixed by anomaly matching. Clearly $\psi_{+2}$ is merely a spectator in the gauge dynamics. The three other fermions are coupled to the $U(1)$ gauge field and, collectively,  flow to a CFT with $c_L=1/2$ and $c_R=0$ which means that there aren't too many options: the theory has to be a single left-moving fermion with charge 1 under the global symmetry $\widehat{G}$. The only remaining question is whether there is, in addition, a TQFT resulting in degenerate ground states. 

\para
Here we show that there is no additional TQFT, and the 3450 theory flows to a single, free Dirac fermion\footnote{This same conclusion was reached using different methods by Philip Boyle Smith.}. 

\para
There is an obvious generalistion of the 3450 model, to gauge theories involving other Pythagorean triples, and we will work with this more general class of theories. The charges under the gauge symmetry $G$ and global symmetry $\widehat{G}$ are given by
\begin{equation}
 	\begin{array}{c|cccc}
&\psi_{-1}&\psi_{-2} & \psi_{+1} & \psi_{+2} \\ \hline
G & \,p^2-q^2\ & 2pq & \ p^2 + q^2 & 0\\
\widehat{G}	& p\ & q & \  p & q
\end{array}\
 \end{equation}
 Here $p$ and $q$ are co-prime integers, and it will be useful to take $q$ odd without loss of generality. This gives the 3450 model when $p=2$ and $q=1$.
 
 \para
 The global symmetry $\widehat{G}$ always acts faithfully. However, the gauge symmetry $G$ is faithful only when $p$ is even, since then $(p^2-q^2 , 2pq ,  p^2 + q^2)$ define a primitive Pythagorean triple. If  $p$ is odd then only a quotient $G/\Z_2$ acts faithfully, and the theory has a $\Z_2$ 1-form symmetry. In either case, the theory is fermionic. 

\para
The dynamics of this gauge theory is again fixed by anomaly matching: it must flow to a purely left-moving fermion, now with charge $q$ under $\widehat{G}$. 
We will study the dynamics of this theory using bosonisation, which will allow us to identify the infra-red TQFT.

\para
It will prove useful to form the Dirac fermions $(\psi_{-1}^\dagger, \psi_{+1})$ and $(\psi_{-2},\psi_{+2}^\dagger)$ and then bosonise to find action
\begin{align}
  S &= \int \left(-\frac{1}{2e^2}F^2+\frac{1}{8\pi}(d \varphi_1-2p^2A-2p\hat{A})^2+\frac{1}{8\pi}(d \varphi_2+2pqA+2q\hat{A})^2\right. \nn\\
   &\left. \hspace{15mm}- \frac{q}{2\pi}\left(q\varphi_1 + p\varphi_2\right) dA   \right) \ ,
\end{align}
where $A$ is a gauge field for $G$, and $\hat{A}$ a background gauge field for $\widehat{G}$. 

\para
There are now a number of fiddly but ultimately trivial manoeuvres to make; let us describe them in words. We first replace $(\varphi_1,\varphi_2)$ with the $2\pi$-periodic scalars $(\hat{\varphi}_1,\hat{\varphi}_2) = (b\varphi_1+a\varphi_2,q\varphi_1 + p\varphi_2)$ where $aq-bp=1$ and we can choose $b$ odd. The virtue of this move is that $A$ couples to the shift current only of $\hat{\varphi}_1$. Thus, by then dualising $\hat{\varphi}_1$ we land on an action in which $A$ couples only to winding currents, which is what we want. One can at this stage make a final linear transformation of fields such that $A$ couples to the winding current of a single field. We finally land on the action
\begin{align}
  S &= \int  \bigg( -\frac{1}{2e^2}F^2+\frac{1}{2\pi}(d\sigma_1)^2 + \frac{4\tilde{b}^2+\tilde{a}^2}{8\pi k} (d \sigma_2)^2 + \frac{2\tilde{b}q+\tilde{a}p}{2\pi k}d\sigma_1 \cdot d \sigma_2 \nn\\
   &\hspace{15mm}\ \ \ \ +\frac{bq+ap}{2\pi k}d\sigma_1\wedge d\sigma_2 - \frac{1}{2\pi} \sigma_2\, dA + \frac{1}{\pi}(q\sigma_1+\tilde{b}\sigma_2)\, d\hat{A}\bigg) \ ,
  \label{eq: cd action}
\end{align}
where $\tilde{a},\tilde{b}$ are some integers satisfying $\tilde{a}q-2\tilde{b}p=1$, which exist since $q$ and $2p$ are co-prime.

\para
We once again have to keep track of how the $\Z_2$ twists change as we make these manoeuvres. The upshot is that the partition function is
\begin{align}
  Z_{p,q} = \frac{1}{4}\sum_{\alpha,\beta,\gamma,\delta}(-1)^{\alpha \beta +\gamma \delta}Z^{(\alpha,\beta;\gamma,\delta)} \ ,
\end{align}
where $Z^{(\alpha,\beta;\gamma,\delta)}$ is the partition function of the action (\ref{eq: cd action}) in which we impose twisted boundary conditions on $\sigma_1$,
\begin{align}
  e^{i\sigma_1(z+\omega_1)} = (-1)^{p\alpha-\gamma} e^{i\sigma_1(z)}\ \ \ {\rm and} \ \ \  e^{i\sigma_1(z+\omega_2)} = (-1)^{p\beta-\delta} e^{i\sigma_1(z)} \ ,
\end{align}
with no twist for $\sigma_2$, and additionally insert the network of topological defects,
  \begin{align}
 \D_{\alpha,\beta,\gamma,\delta} = &\exp\left[-\frac{i}{2} \Big(2p\gamma \int_{\omega_2}d\sigma_1+\left(p\alpha + \alpha  -\gamma\right)\int_{\omega_2}d\sigma_2\Big)\right]\nn\\
 &\times \exp\left[-\frac{i}{2} \Big(2p\delta \int_{\omega_1}d\sigma_1+\left(p\beta + \beta -\delta\right)\int_{\omega_1}d\sigma_2\Big)\right] \ .
\end{align}
Integrating out $A$ sets $\sigma_2=0$ (mod $2\pi$) which is an allowed configuration in every sector, as $\sigma_2$ does not have twisted boundary conditions. The key new feature of this theory is that even once we set $\sigma_2=0$, the defect does not become trivial precisely because of the twisted boundary conditions for $\sigma_1$. Indeed, we have
 \be
 \D_{\alpha,\beta,\gamma,\delta} &=& \exp\left[-\frac{i}{2} \Big(2p\gamma \int_{\omega_2}d\sigma_1\Big)\right]\exp\left[-\frac{i}{2} \Big(2p\delta \int_{\omega_1}d\sigma_1\Big)\right] \nn\\ &=& (-1)^{p\delta(\alpha-\gamma)+p\gamma(\beta-\delta)} \ .
\ee
Thus, after integrating out $A$ we find the action
\begin{align}
  S &= \int \left( \frac{1}{2\pi}(d\sigma_1)^2+ \frac{2q}{2\pi}\sigma_1\, d\hat{A}\right) \ ,
\label{eq: chiral final action}
\end{align}
and the partition function
\begin{align}
  Z_{p,q} &= \frac{1}{4}\sum_{\alpha,\beta,\gamma,\delta}(-1)^{\alpha \beta +\gamma \delta+p\gamma(\alpha-1)+p\delta(\beta-1)}Z_\text{s}^{(p\alpha-\gamma,p\beta-\delta)}	\ .
\end{align}
If $p$ is even, so the UV theory has no 1-form symmetry, then we find
\begin{align}
  Z_{p,q} = \frac{1}{2} \sum_{\alpha,\beta}(-1)^{\alpha\beta}Z_\text{s}^{(\alpha,\beta)} \ .
\end{align}
Looking at the action (\ref{eq: chiral final action}) and using (\ref{eq: current matching}), we identify this as precisely the partition function of a single Dirac fermion of axial charge $q$ under the $U(1)$ global symmetry $\widehat{G}$, which is of course equivalent to a single Dirac fermion with vector charge $q$ under $\widehat{G}$. There is no TQFT.

\para
Conversely, if $p$ is odd and the UV theory has a $\Z_2$ 1-form symmetry, we find
\begin{align}
  Z_{p,q} = 2\times \left(\frac{1}{2} \sum_{\alpha,\beta}(-1)^{\alpha\beta}Z_\text{s}^{(\alpha,\beta)}\right) \ .
\end{align}
Thus, as we might have predicted, we flow in the IR to a single Dirac fermion of charge $q$ under $\widehat{G}$, along with a TQFT with two vacua.

\subsection{Two Fermions, Two Gauge Fields}\label{22sec}

For our next example, we consider two Dirac fermions coupled to two gauge fields. Each gauge field, individually, will couple in a vector-like manner. But, combined, this can result in a chiral theory.

\para
To see this, first take a $U(1)$ gauge symmetry, which we call $G_1$, coupled to two Dirac fermions with charges $p$ and $q$. 
 What other $U(1)$ symmetries can we also gauge?

\para
We write the left-moving fermions as $\psi_{-1}$ and $\psi_{-2}$, and the right-moving fermions as $\psi_{+1}$ and $\psi_{+2}$. There are two options for the second gauge group, consistent with anomalies. They are:
 \be
\begin{array}{c|cccc}
&\psi_{-1}&\psi_{-2} & \psi_{+1} & \psi_{+2} \\ \hline
G_1 & p & q &  p & q \\
G_2 & q & -p &  q & -p   \\
\end{array}\ \ \ \ \ \  {\rm and}\ \ \ \ \ \  
\begin{array}{c|cccc}
&\psi_{-1}&\psi_{-2} & \psi_{+1} & \psi_{+2} \\ \hline
G_1 & p & q &  p & q \\
G_2 & q & -p &  -q & p   \\
\end{array}
\label{gf1}\ee
The first of these is a vector-like theory, in the sense that we can always add mass terms $\psi_{+1}^\dagger\psi_{-1}$ and $\psi_{+2}^\dagger\psi_{-2}$ consistent with both gauge symmetries. But the second is a chiral theory, with mass terms forbidden, even though individually $G_1$ and $G_2$ are each vector-like; it's only the combination of the two that forbids the mass term. Finally note that both theories are bosonic, in the sense that $(-1)^F$ is contained in $G_1\times G_2$, for all co-prime $p,q$. We will now deal with each of these in turn.

\subsubsection*{The Vector-Like Theory}

We start with the first, vector-like, set of charges in \eqn{gf1}. Define 
\be k=p^2 + q^2\ .\ee
This theory has a $\Z_k$ 1-form symmetry, since the subgroup of the gauge group generated by $(e^{2\pi i p/k},e^{2\pi i q/k})\in G_1\times G_2$ acts trivially. There is correspondingly a $\Z_k$ non-anomalous 0-form symmetry which we can take to be generated by $(\psi_{-1},\psi_{-2},\psi_{+1},\psi_{+2})  \to  (e^{2\pi ip/k}\psi_{-1},e^{2\pi i q/k}\psi_{-2},\psi_{+1},\psi_{+2})$.

\para 
We will introduce a  gauge fields $A_1,A_2$ for the symmetries $G_1,G_2$, respectively. Now both $G_1$ and $G_2$ are gauged so both $A_1$ and $A_2$ are dynamical. We  pair  the fermions into  Dirac fermions $(\psi_{-1},\psi_{+1})$ and $(\psi_{-2},\psi_{+2})$ and then bosonise to find action
\be
  S &=& \int \bigg(-\frac{1}{2e_1^2}(F_1)^2 - \frac{1}{2e_2^2}(F_2)^2+\frac{1}{8\pi}(d \varphi_1)^2+\frac{1}{8\pi}(d \varphi_2)^2\nn\\
  &&\ \ \ \ \ \ \ \ \ \ \ \ -\ \frac{1}{2\pi}\left(p\varphi_1 + q\varphi_2\right) dA_1 - \frac{1}{2\pi}\left(q\varphi_1 -p\varphi_2\right) dA_2 \bigg)
\ .\ee
Integrating out $A_1$ gaps out the combination $(p\varphi_1+q\varphi_2)$ with a single vacuum, leaving behind an action for the other combination $\sigma=(a\varphi_1+b\varphi_2)$ with $aq-bp=1$,
\begin{align}
  S = \int  \left( -\frac{1}{2e_2^2}(F_2)^2+\frac{k}{8\pi}(d \sigma)^2  - \frac{k}{2\pi}\sigma\, dA_2 \right) \ .
\end{align}
Now integrating out $A_2$ gaps $\sigma$, but now with $k$ vacua. Furthermore everything we've done flies just the same in all 15 $\Z_2$  twisted sectors, as we are always just twisting in winding symmetries. We thus arrive at low energies at the partition function
\begin{align}
  Z_{p,q} = k\times \left(\frac{1}{4}\sum_{\alpha,\beta,\gamma,\delta}(-1)^{\alpha\beta + \gamma\delta}\right) = k\ .
\end{align} 
Thus, the theory leaves behind a $\Z_k$ gauge theory, corresponding to the spontaneously broken $\Z_k$ 0-form symmetry.

\subsubsection*{The Chiral Theory}

Now let's look at the second charge assignment in \eqn{gf1}. This is a chiral theory. There is then a qualitative distinction to be made depending on the parity of the product $pq$.

\para
If $pq$ is even, so that one of $p,q$ is even, then the gauge group acts faithfully. Accordingly there is no 1-form global symmetry, and the theory also has no non-anomalous global symmetries.

\para
Conversely if $pq$ is odd, then the $\Z_2$ generated by $(-1,-1)\in G_1\times G_2$ acts trivially, corresponding to a $\Z_2$ 1-form symmetry. Correspondingly in this case there is a $\Z_2$ non-anomalous global 0-form symmetry, which we can take to act as $(\psi_{-1},\psi_{-2},\psi_{+1},\psi_{+2}) \to (-\psi_{-1},-\psi_{-2},\psi_{+1},\psi_{+2})$.

\para
We treat the two cases together and consider generic co-prime $p,q$. After bosonisation we have the action
\be
  S &=&\int \ \bigg(\, -\frac{1}{2e_1^2}(F_1)^2 - \frac{1}{2e_2^2}(F_2)^2+\frac{1}{8\pi}(d \varphi_1+2qA_2)^2+\frac{1}{8\pi}(d \varphi_2-2pA_2)^2\nn\\
   &&\ \ \ \ \ \ \ \ \  \ \ \ -\ \frac{1}{2\pi}\left(p\varphi_1 + q\varphi_2\right) dA_1 \bigg) \ .
  \label{eq: actionactionaction}
\ee
Integrating out $A_1$ gaps the combination $(p\varphi_1+q\varphi_2)$ with a single vacuum, leaving behind an action for the other combination $\sigma=(a\varphi_1+b\varphi_2)$ with $aq-bp=1$,
\begin{align}
  S = \int \left(-\frac{1}{2e_2^2}(F_2)^2+\frac{k}{8\pi}(d \sigma+2A_2)^2\right) \ .
\end{align}
Dualising, we have
\begin{align}
  S = \int   \left(-\frac{1}{2e_2^2}(F_2)^2+\frac{1}{2\pi k}(d \tilde{\sigma})^2 + \frac{1}{\pi}\tilde{\sigma}\, dA_2\right)\ .
\end{align}
Then integrating out $A_2$, we gap out $\tilde{\sigma}$ with two vacua. Thus, the compact boson theory (\ref{eq: actionactionaction}) has two ground states.
\para 
Next, we must sum over all 16 topological sectors to get the right answer for the theory of fermions. Recall that the partition function of the UV theory is given by
\begin{align}
  Z_{p,q} = \frac{1}{4}\sum_{\alpha,\beta,\gamma,\delta}(-1)^{\alpha\beta + \gamma\delta}Z_\text{w}^{(\alpha,\beta;\gamma,\delta)}\ .
\end{align}
Following similar steps as in previous sections, we find as we go to low energies,
\begin{align}
  Z_\text{w}^{(\alpha,\beta;\gamma,\delta)} \to 2\delta_{q\alpha - p \gamma,0}\delta_{q\beta-p\delta,0} \ ,
\end{align}
where the Kronecker symbols are understood modulo $2$. Thus, at low energies we find the partition function
\begin{align}
  Z_{p,q} &\to  \frac{1}{2}\sum_{\alpha,\beta,\gamma,\delta}(-1)^{\alpha\beta + \gamma\delta}\delta_{q\alpha - p \gamma,0}\delta_{q\beta-p\delta,0}		= \left\{\begin{aligned}
  	\,\,1 \qquad &\text{if }pq \text{ even}		\\
  	\,\,2 \qquad &\text{if }pq \text{ odd}
  \end{aligned} \right. \ .
\end{align}
Thus we see that the theory has a unique ground state only in the case $pq$ even. Otherwise, it has a two-fold degeneracy of ground states, corresponding to the spontaneously broken $\Z_2$ global 0-form symmetry. 

\subsection{The Higgs Phase and Symmetric Mass Generation}\label{smgsec}

Symmetric mass generation is the name given to any mechanism that gaps a theory while preserving a non-anomalous chiral symmetry.  Here we will look more closely at a mechanism to gap a pair of chiral fermions in 2d. 

\para
The simplest example is the 3450 model. We consider free fermions with charges as in \eqn{3450}, but now where both $G$ and $\hat{G}$ are both viewed as global symmetries (i.e. there is no dynamical gauge field). 
The challenge of symmetric mass generation is to find a way to deform the theory of free fermions so that the system becomes gapped {\it without} breaking the $U(1)$ symmetry $G$. (You could also require that the global symmetry $\widehat{G}$ in \eqn{3450} is unbroken.)

\para
Various, related, methods to affect symmetric mass generation have been proposed in the literature. There is particular interest in performing symmetric mass generation on the lattice as it promises an avenue to constructing discrete versions of chiral gauge theories. In that context, it was suggested that symmetric mass generation could could be induced by turning on certain irrelevant operators, with a coefficient that is comparable to lattice scale \cite{xwen,jwen}, and attempts to implement this  proposal in  DMRG simulations have been made  
\cite{3450,fail}.

\para
In the field theoretic context, a more palatable approach, albeit one that is restricted in its utility to two dimensions, is to make use of the marginal current-current interactions enjoyed by free fermions. As one moves in the moduli space of conformal field theories, the dimension of the (initially) irrelevant operator changes until we reach a point where they become relevant. The idea is that these operators then induce an RG flow that results in symmetric mass generation. 

\para
It was pointed out in \cite{mesmg} that these different approaches could be viewed as the result of  2d gauge dynamics.  The idea, which we will develop more fully below,  is to look at the dynamics of 2d gauge theories in the Higgs phase. If the gauge theory is coupled to sufficient amount of matter (both bosons and fermions) then it's possible to realise the free fermions, with chiral charges under a global $U(1)$ symmetry, in the infra-red. Changing the vacuum expectation value of the condensed scalars then changes the dimension of the (initially) irrelevant operators until they condense.

\para
This story has features in common with the gauge theory described in Section \ref{scalarfermionsec}, where we saw that changing the Higgs \vev moves us along the line of $c=1$ fixed points. However, in that context we didn't have a massless {\it phase}, because we had to fine-tune the fermion mass $m_f$. The purpose of this section is to explore the same set-up, but within the context of a chiral gauge theory. We will see that, again, we move in the space of CFTs, but without the need to fine-tune any mass parameter in the UV.

\para
The simplest model of symmetric mass generation takes the chiral gauge theory on the right-hand-side of \eqn{gf1} and adds two scalars, $\phi_1$ and $\phi_2$, introducing the possibility of a Higgs phase. The matter content is
\be
\begin{array}{c|cccc:cc}
&\psi_{-1}&\psi_{-2} & \psi_{+1} & \psi_{+2}  & \phi_1 & \phi_2\\ \hline
G_1 & p & q &  p & q & 1 & 0  \\
G_2 & q & -p &  -q & p  & 0 & 1 \\ 
\widehat{G}_1 & 0 & 0 &  0 & 0 & -1 & 0 \\
\widehat{G}_2 & 0 & 0 & 0 & 0 & 0 & -1
\end{array}
\label{eq: SMG theory}
\ee
Here $G_1,G_2$ are gauge symmetries while $\widehat{G}_1,\widehat{G}_2$ are global symmetries. Requiring these two global symmetries prohibits Yukawa terms in the theory. Note the addition of the scalars means that $(-1)^F$ is not gauged, and the theory is fermionic.

\para 
We specialise to the case that $p,q$ are co-prime with $pq$ even. If we give a mass to the newly-added scalars $\phi_1,\phi_2$, then the results of Section \ref{22sec} mean that we flow to a gapped phase with a unique ground state. This means that getting the gapped phase is trivial in this set-up: it is the gapless phase that will prove to be more subtle.

\para
We expect the gapless phase to arise when we condense the scalars. If we can trust the classical analysis, then, when the scalars are condensed, the global symmetries get twisted by the gauge symmetries, so that the diagonal subgroup of  $G_i\times \widehat{G}_i$ is unbroken for $i=1,2$. In this way, the chiral charge assignments of the fermions under $G_i$ become global charge assignments. We would then be in a situation where we have massless fermions in the Higgs phase, carrying chiral charges under a global symmetry, and a fully gapped phase in which that same global symmetry is unbroken. This would be a successful  symmetric mass generation. 

\para
The question is: can we trust the classical analysis in the Higgs phase? It's not  immediately obvious. As we've seen in Section \ref{scalarfermionsec}, a $U(1)$ gauge field coupled to a single Dirac fermion and a scalar is typically gapped, even when the scalar is condensed. In that context, we could fine-tune to a critical point (actually a critical line) by dialing the fermion mass $m_f$. To implement symmetric mass generation, it's important that we have a gapless phase, without any fine-tuning.  We will now show that, happily, is indeed the case. Moreover, we will get a handle on the phase diagram, understanding what operators become relevant and gap the system as we dial the scalar masses to move from the Higgs to confining phase. 

\para
We condense the scalars, and write $\phi_i = v_i e^{i\sigma_i}$. Deep in the  Higgs phase, we have the action
\be
  S &=& \int  \bigg( - \frac{1}{2e_1^2}(F_1)^2- \frac{1}{2e_2^2}(F_2)^2+\frac{1}{8\pi}(d \varphi_1+2qA_2)^2+\frac{1}{8\pi}(d \varphi_2-2pA_2)^2 \nn\\
  &&\hspace{20mm}+\ v_1^2(d \sigma_1+ A_1 - \hat{A}_1)^2+ v_2^2(d \sigma_2+ A_2 - \hat{A}_2)^2 - \frac{1}{2\pi}\left(p\varphi_1 + q\varphi_2\right) dA_1 \bigg) \ ,\nn\\
\label{eq: SMG Higgs phase}
\ee
where $A_1,A_2$ are dynamical gauge fields for $G_1,G_2$, respectively. We have also turned  on background gauge fields $\hat{A}_1,\hat{A}_2$ for $\widehat{G}_1,\widehat{G}_2$, respectively, to help us keep track of charges under the global symmetry. As usual, we will be blas\'e about $\Z_2$ quotients for the time being and just consider the compact boson theory to start with.

\para
We next dualise the two scalars $\sigma_1,\sigma_2$ to find
\be
  S &=& \int \bigg(-\frac{1}{2e_1^2}(F_1)^2 - \frac{1}{2e_2^2}(F_2)^2+\frac{1}{8\pi}(d \varphi_1+2qA_2)^2+\frac{1}{8\pi}(d \varphi_2-2pA_2)^2 \nn\\
  &&\hspace{20mm}+\ \frac{1}{16\pi^2 v_1^2}(d \tilde{\sigma}_1)^2+ \frac{1}{16\pi^2 v_2^2}(d \tilde{\sigma}_2)^2 
  \nn\\
  &&\hspace{20mm}- \ \frac{1}{2\pi}\left(p\varphi_1 + q\varphi_2-\tilde{\sigma}_1\right) dA_1+\frac{1}{2\pi}\tilde{\sigma}_2 dA_2 - \frac{1}{2\pi}\left( \tilde{\sigma}_1d\hat{A}_1  + \tilde{\sigma}_2d\hat{A}_2\right)\bigg)\ . \nn
\ee
Integrating out $A_1$ gaps the combination $(p\varphi_1 + q\varphi_2 - \tilde{\sigma}_1)$. We introduce 
 the $2\pi$-periodic scalar $\rho = a\varphi_1 + b\varphi_2$ with $aq-bp=1$, and get the action
\be
  \hspace{-1cm} S = \int  \bigg(- \frac{1}{2e_2^2}(F_2)^2+\frac{1}{8\pi}(q d \rho-b d\tilde{\sigma}_1+2qA_2)^2+\frac{1}{8\pi}(-pd \rho + ad\tilde{\sigma}_1-2pA_2)^2 \bigg) \ .
  \ee
Finally, we want to integrate out $A_2$. To get there, we dualise $\rho$ to find
\begin{align}
  S &= \int \bigg( -\frac{1}{2e_2^2}(F_2)^2+\frac{1}{2\pi k}(d \tilde{\rho})^2 + \frac{1}{8\pi k}\left(1+\frac{k}{2\pi v_1^2}\right)(d \tilde{\sigma}_1)^2 + \frac{1}{16\pi^2 v_2^2}(d \tilde{\sigma}_2)^2  \nn\\
  &\hspace{15mm}- \frac{ap+bq}{2\pi k} d\tilde{\sigma}_1 \wedge d\tilde{\rho}+\frac{1}{2\pi}\left(\tilde{\sigma}_2+2 \tilde{\rho} \right) dA_2 - \frac{1}{2\pi}\left( \tilde{\sigma}_1 d\hat{A}_1  + \tilde{\sigma}_2d\hat{A}_2\right)\bigg) \ .
\end{align}
Integrating out $A_2$ gaps the combination $(\tilde{\sigma}_2 + 2\tilde{\rho})$. After performing a final duality transformation on the resulting action, we arrive at our final action
\begin{align}
  S &= \int \bigg(\frac{1}{8\pi k}\left(1+\frac{k}{2\pi v_1^2}\right)(p D\varphi_1+qD\varphi_2)^2 + \frac{1}{8\pi k}\left(1+\frac{k}{2\pi v_2^2}\right)^{-1}\left(q D\varphi_1 - pD\varphi_2\right)^2\nn\\
  &\hspace{25mm} -\ \frac{1}{2\pi} \hat{A}_1 \wedge \left(pD\varphi_1 +qD\varphi_2\right) \bigg) \ ,
  \label{eq: Higgs final action}
\end{align}
where $\varphi_1,\varphi_2$ are both  $2\pi$-periodic scalars, with covariant derivatives
\begin{align}
  D\varphi_1 = \partial \varphi_1 + 2q\hat{A}_2\ \ \ {\rm and}\ \ \  D\varphi_2 = \partial \varphi_2 - 2p\hat{A}_2 \ .
\end{align}
Following carefully the fate of the $\Z_2$ twists, we learn that the low energy partition function is simply
\begin{align}
  Z_{p,q} = \frac{1}{4}\sum_{\alpha,\beta,\gamma,\delta}(-1)^{\alpha\beta + \gamma\delta} Z_\text{w}^{(\alpha,\beta;\gamma,\delta)} \ ,
\label{eq: c=2 twist}
\end{align}
where as in pervious sections, this notation means we twist by $(\alpha,\beta)$ in the $U(1)_\text{winding}$ of $\varphi_1$, and a twist by $(\gamma,\delta)$ in the $U(1)_\text{winding}$ of $\varphi_2$.

\para
First note that, asymptotically as $v_1^2,v_2^2\to \infty$, the action \eqn{eq: Higgs final action} coincides with our earlier  (\ref{eq: actionactionaction}), but with the dynamical gauge fields in \eqn{eq: actionactionaction} replaced by background gauge fields in \eqn{eq: Higgs final action}. This is telling us that, deep in the Higgs phase, the theory is a collection of massless fermions coupled in a chiral fashion to backgound global symmetries with charges
 \be
\begin{array}{c|cccc}
&\psi_{-1}&\psi_{-2} & \psi_{+1} & \psi_{+2} \\ \hline
\widehat{G}_1 & p & q &  p & q \\
\widehat{G}_2 & q & -p &  -q & p   \\
\end{array}
\ee
In particular, there is a  subgroup $\hat{H}\subset \widehat{G}_1\times \widehat{G}_2$ with charges
 \be
\begin{array}{c|cccc}
&\psi_{-1}&\psi_{-2} & \psi_{+1} & \psi_{+2} \\ \hline
\hat{H} & p^2-q^2 & 2pq &  p^2+q^2 & 0 
\end{array}
\ee
For the case of  $p=2,q=1$, this is the 3450 global symmetry.
\para 
The action (\ref{eq: Higgs final action}) tells us how we move on the $c=2$ conformal manifold as we decrease $v_1^2, v_2^2$, at least to leading order in $1/v_1^2$ and $1/v_2^2$. This conformal manifold is four (real) dimensional, with 3 moduli contained in the metric on the target space torus, and the fourth coming from a $B$-field. We thus find that, again at leading order, we trace out a 2-dimensional subspace of this conformal manifold. Indeed, we can write 
\begin{align}
  S &= \int \bigg( \frac{G_{ij}}{4\pi} D\varphi_i \cdot D\varphi_j- \frac{1}{2\pi} \hat{A}_1 \wedge \left(pD\varphi_1 +qD\varphi_2\right) \bigg) \ ,
  \label{c2}
    \end{align}
with
\begin{equation}
\begin{aligned}
  G_{11} 	&= \frac{1}{2 k} \left[\left(1+\frac{k}{2\pi v_1^2}\right)p^2 + \left(1+\frac{k}{2\pi v_2^2}\right)^{-1}q^2 \right] &&\hspace{-15mm}		\\
  G_{22}	&= \frac{1}{2 k} \left[\left(1+\frac{k}{2\pi v_1^2}\right)q^2 + \left(1+\frac{k}{2\pi v_2^2}\right)^{-1}p^2 \right]	&&\hspace{-15mm}	\\
  G_{12} = G_{21}	&= \frac{pq}{2 k}\left[\left(1+\frac{k}{2\pi v_1^2}\right) - \left(1+\frac{k}{2\pi v_2^2}\right)^{-1}\right] \ , &&\hspace{-15mm} \label{eq:Gmat}
\end{aligned}
\end{equation}
which does indeed approach $G_{ij} = \frac{1}{2}\delta_{ij}$ in the limit $v_1^2,v_2^2\to \infty$. 

%


\subsubsection*{Stability and Phase Diagram}

\para
We have seen that when $m_1^2,m_2^2$ are both large and positive the theory is trivially gapped, while if they are both large and negative, we find the $c=2$ theory (\ref{c2}) which for $v_1^2,v_2^2\to \infty$ just describes a pair of free fermions. 

\para
In the Higgs phase, as we vary the expectation values $v_i$, $i=1,2$, we move in the $c=2$ moduli space and the spectrum of the theory varies. We would like to understand when, and how, the $c=2$ phase becomes unstable due to the operators singlets under the global symmetry becoming relevant. 
Our gauge theory is strongly coupled, which means that we will not be able to pinpoint the exact values of the masses where we exit the $c=2$ phase. Nonetheless, we will identify which operators are most likely to cross marginality first, and destabilise the CFT.

\para
Before we get there, there is still more we can learn about the asymptotic phase structure: what happens in the other two corners of the phase diagram?

\para
If we give a large positive mass to $\phi_1$ while condensing $\phi_2$, then we Higgs $G_2$ and end up with the $(p,q)$ 2-flavour massless Schwinger model, which in turn flows to a $c=1$ fermionic CFT of radius $R^2 = k/2$. By following similar (and indeed simpler) steps to those earlier in this section, we can refine this picture to incorporate the leading effect of a large but finite \vev $v_2$.  We find that the radius is corrected to
\begin{align}
    R^2 = \frac{k}{2}\left( 1+\frac{k}{2\pi v_2^2} \right)^{-1} \ .
\label{eq: c=1 radius}
\end{align}
The global symmetry $\widehat{G}_1$ acts trivially on this theory, while $\widehat{G}_2$ coincides precisely with the axial symmetry. The $\widehat{G}_2$ invariant scalar operators $\cos(2n\tilde{\varphi})$ have dimension $\Delta = 2R^2 n^2$, and so the far region $v_2^2\to \infty$ is indeed stable. As $v_2^2$ is reduced, the operator $\cos(2\tilde{\varphi})$ is generated and trivially gaps\footnote{Naively this potential has two ground states $\tilde{\varphi}=0,\pi$. But these are mapped into each other by the $\Z_2$ twist; a careful partition function analysis similar to as in previous sections shows that this term does indeed trivially gap the theory.} the theory.

\para
The analysis is identical in the opposite corner, where we gap $\phi_2$ while condensing $\phi_1$.  We find again a $c=1$ fermionic CFT now with 
\begin{align}
    R^2 = \frac{k}{2}\left( 1+\frac{k}{2\pi v_1^2} \right)^{-1} \ .
\label{eq: c=1 radius 2}
\end{align}
The crucial difference is that now this theory is inert under $\widehat{G}_2$, while $\widehat{G}_1$ coincides with the axial symmetry. Because the gapless excitations are charged under different global symmetries, the two $c=1$ asymptotic regions cannot be continuously connected. So what happens in the interior of the phase diagram? To answer this we return to the $c=2$ theory (\ref{c2}).
\para
The charge-conjugation invariant vertex operators of the $c=2$  theory \eqref{c2} can be written as \cite{Dulat:2000xj,Damia:2024xju}
\begin{equation} \label{eq:vertex_op_bos}
    \hat{\cal O}_{\mathbf{n},\mathbf{m}} = \cos\left( \mathbf{n} \cdot \boldsymbol{\varphi}+\mathbf{m} \cdot \boldsymbol{\tvp}\right) = \hat{\cal O}_{-\mathbf{n},-\mathbf{m}} \ ,
\end{equation}
 where we defined the vectors $\boldsymbol{\varphi} = (\varphi_1,\varphi_2)$ and $\boldsymbol{\tvp} = (\tvp_1,\tvp_2)$, as well as $\mathbf{n} = (n_1,n_2)$ and $\mathbf{m} = (m_1,m_2) $. The twisting (\ref{eq: c=2 twist}) tells us that the local operators of the theory are those with $m_1,m_2\in \Z$ and $n_1+(m_1/2), n_2 + (m_2/2)\in \Z$.
 The scaling dimension and spin of these operators in the theory \eqref{c2} are 
 \begin{equation}
     \begin{aligned}
         \Delta_{\mathbf{n},\mathbf{m}} &=h_{\mathbf{n},\mathbf{m}}+\bar{h}_{\mathbf{n},\mathbf{m}}= \frac{1}{2}\, \mathbf{n}\cdot G^{-1}\cdot \mathbf{n}+\frac{1}{2}\, \mathbf{m}\cdot G\cdot \mathbf{m}\\
         s_{\mathbf{n},\mathbf{m}} &= |h_{\mathbf{n},\mathbf{m}}-\bar{h}_{\mathbf{n},\mathbf{m}}|=|\mathbf{n} \cdot \mathbf{m}| \ ,
     \end{aligned}
 \label{eq: c=2 spectrum}
 \end{equation}
 with the matrix $G$ defined in \eqref{eq:Gmat}.

\para
We are interested in operators that are invariant under the global symmetries $\widehat{G}_1\times \widehat{G}_2$. These are precisely the operators $\mathcal{O}_{r,s} = \mathcal{O}_{-r,-s} =  \hat{\mathcal{O}}_{r(p,q),2s(q,-p)}$ for integers $r,s$. All such operators are Lorentz scalars. 
Their  scaling dimensions are
  \begin{equation}
      \Delta_{r,s} = k\left[r^2 \left(1+\frac{k}{2\pi v_1^2} \right)^{-1} +s^2 \left(1+\frac{k}{2\pi v_2^2} \right)^{-1}\right] \ .  \label{eq:Vrs}
  \end{equation}
%
The computation \eqref{c2} holds deep in the Higgs phase, and, for $v_i^2 \to \infty$, we find operators with dimensions $\Delta = k (r^2 + s^2)$. Since $k\ge 5$, we see that deep in the Higgs phase there are no relevant deformations compatible with the symmetry, which ensures stability of the fixed point for large enough $v_i^2$'s.

\para
Deep in Higgs phase, the lowest dimensional operators scalar singlets have dimension
\begin{equation}
\Delta_{1,0} = k \left(1+\frac{k}{2\pi v_1^2}\right)^{-1}\ \ \ {\rm and}\ \ \  \Delta_{0,1} = k  \left(1+\frac{k}{2\pi v_2^2}\right)^{-1}\,. \label{eq:lowest_dims}
\end{equation}
%
As we move out of the Higgs phase by lowering the value of $v_i^2$, the dimension of these operators decreases, until one of them crosses marginality and becomes relevant. At this point the $c=2$ CFT becomes unstable: we do not have other parameters to tune this relevant operator away, which will then be generated and start a flow from the $c=2$ theory. From \eqref{eq:lowest_dims}, we would expect that the region where the $c=2$ theory becomes unstable sits at $v_i^2 = {k}/(\pi({k-2}))$. Clearly, corrections which can be neglected deep in the Higgs phase will become more and more important as we decrease $v_i^2$, which makes it impossible to precisely identify the region of stability of the $c=2$ theory. However, we do not expect these corrections to change the picture qualitatively.

\para
We would like to understand where the theory is driven to when these operators become relevant. The current formulation (\ref{c2}) obscures this somewhat, since the $\mathcal{O}_{r,s}$ are written in terms of not only $\varphi_1,\varphi_2$ but also their duals $\tilde{\varphi}_1,\tilde{\varphi}_2$, and thus are not local deformations of the Lagrangian. To make progress we move to an alternative formulation.

\para
We first make the field redefiniton
\begin{equation} \label{eq:changeofvars}
    \boldsymbol{\varphi} = M \cdot \boldsymbol{\rho} \ \ \ {\rm with}\ \ \  M  = \begin{pmatrix}
        -b & q\\
        a & -p
    \end{pmatrix} \ ,
\end{equation}
with $aq-bp=1$. We then dualise $\rho_2$. The resulting action is
\begin{align}
  S &= \int \bigg(\frac{1}{8\pi k}\left(1+\frac{k}{2\pi v_1^2}\right)(d \rho_1)^2+\frac{1}{2\pi k}\left(1+\frac{k}{2\pi v_2^2}\right)(d \tilde{\rho}_2)^2 \nn\\
  &\hspace{15mm}-\frac{ap+bq}{2\pi k} d\rho_1 \wedge d\tilde{\rho}_2  - \frac{1}{2\pi}\left( \rho_1 d\hat{A}_1  -2 \tilde{\rho}_2 d\hat{A}_2\right)\bigg) \ .
\label{c3}
\end{align}
We can follow the fate of the $\Z_2$ gauging (\ref{eq: c=2 twist}) as we perform these manipulations. We can always choose one of $(a,b)$ to be even\footnote{We can alternatively take both $(a,b)$ odd, which results in a slightly different partition function. This reflects the non-invariance of $e^{iS}$ under $(a,b)\to (a+p,b+q)$ due to sectors in which $\tilde{\rho}_2$ has half-integer winding. One can nonetheless proceed with such a choice of $(a,b)$, and the conclusions are unchanged.}, in which case the partition function is
\begin{align}
  Z_{p,q} = \frac{1}{4}\sum_{\alpha,\beta,\gamma,\delta}(-1)^{\alpha\beta + \gamma\delta} Z^{(\alpha,\beta;\gamma,\delta)} \ ,
\end{align}
where $ Z^{(\alpha,\beta;\gamma,\delta)}$ is computed with a $(\alpha,\beta)$ twist in the $U(1)_\text{winding}$ of $\rho_1$, and a $(\gamma,\delta)$ twist in the $U(1)_\text{shift}$ of $\tilde{\rho}_2$. We see that two such choices for $(a,b)$ are related by $(a,b)\to (a+2p,b+2q)$, under which $e^{iS}$ is invariant.

\para
This action has a number of benefits versus (\ref{c2}). The metric is now diagonal, albeit at the expense of introducing a B-field coupling. For our purposes the key feature is that $\widehat{G}_1$ couples only to the winding current of $\rho_1$, while $\widehat{G}_2$ couples only to the winding current of $\tilde{\rho}_2$. Indeed, our $\widehat{G}_1\times \widehat{G}_2$ invariant operators take the simple form
\begin{align}
  \mathcal{O}_{r,s} = \cos\left(r\rho_1 + 2s \tilde{\rho}_2 \right) \ ,
\end{align}
which are indeed local with respect to the fields of the Lagrangian (\ref{c3}).

\para
Now let's consider what happens when we vary one of the VEVs, says $v_1^2$, while keeping the other VEV $v_2^2$ large and fixed. When we reach the point where the operator $\mathcal{O}_{1,0}$ becomes relevant, an RG flow will drive us away from the $c=2$ fixed point. Indeed, the addition of the term $\cos(\rho_1)$ to the action (with either sign) gaps $\rho_1$ with a unique vacuum. We land on the action
\begin{align}
  S &= \int \bigg(\frac{1}{2\pi k}\left(1+\frac{k}{2\pi v_2^2}\right)(d \tilde{\rho}_2)^2  + \frac{1}{\pi}  \tilde{\rho}_2 d\hat{A}_2 \bigg) \ .
\end{align}
After a T-duality, and keeping track of the necessary $\Z_2$ twists, we determine the fixed point of the flow as a $c=1$ fermionic CFT with radius given in (\ref{eq: c=1 radius}), where $\widehat{G}_1$ acts trivially while $\hat{A}_2$ couples with charge $1$ to the axial current. This is indeed the theory we found in the asymptotic regime $m_1^2\gg e^2$, $v_2^2\gg 1$, strongly suggesting that there is no further phase transition as we decrease $v_1^2$.

\para
The opposite direction in the phase diagram works a little differently. Fixing $v_1^2$ and reducing $v_2^2$, we reach a point where $\mathcal{O}_{0,1}=\cos(2\tilde{\rho}_2)$ becomes relevant, gapping out $\tilde{\rho}$ with a single ground state by virtue of the gauged $\Z_2$ shift symmetry of $\tilde{\rho}_2$. We land on the action
\begin{align}
  S &= \int \bigg(\frac{1}{8\pi k}\left(1+\frac{k}{2\pi v_1^2}\right)(d \rho_1)^2  - \frac{1}{2\pi} \rho_1 d\hat{A}_1 \bigg) \ .
\end{align}
Accounting for the necessary $\Z_2$ twist, we land on a $c=1$ fermionc CFT with radius
\begin{align}
R^2 = \frac{1}{2k} \left(1+\frac{k}{2\pi v_1^2}\right) \ ,
\end{align}

\begin{center}
\centering
\includegraphics[width=50mm]{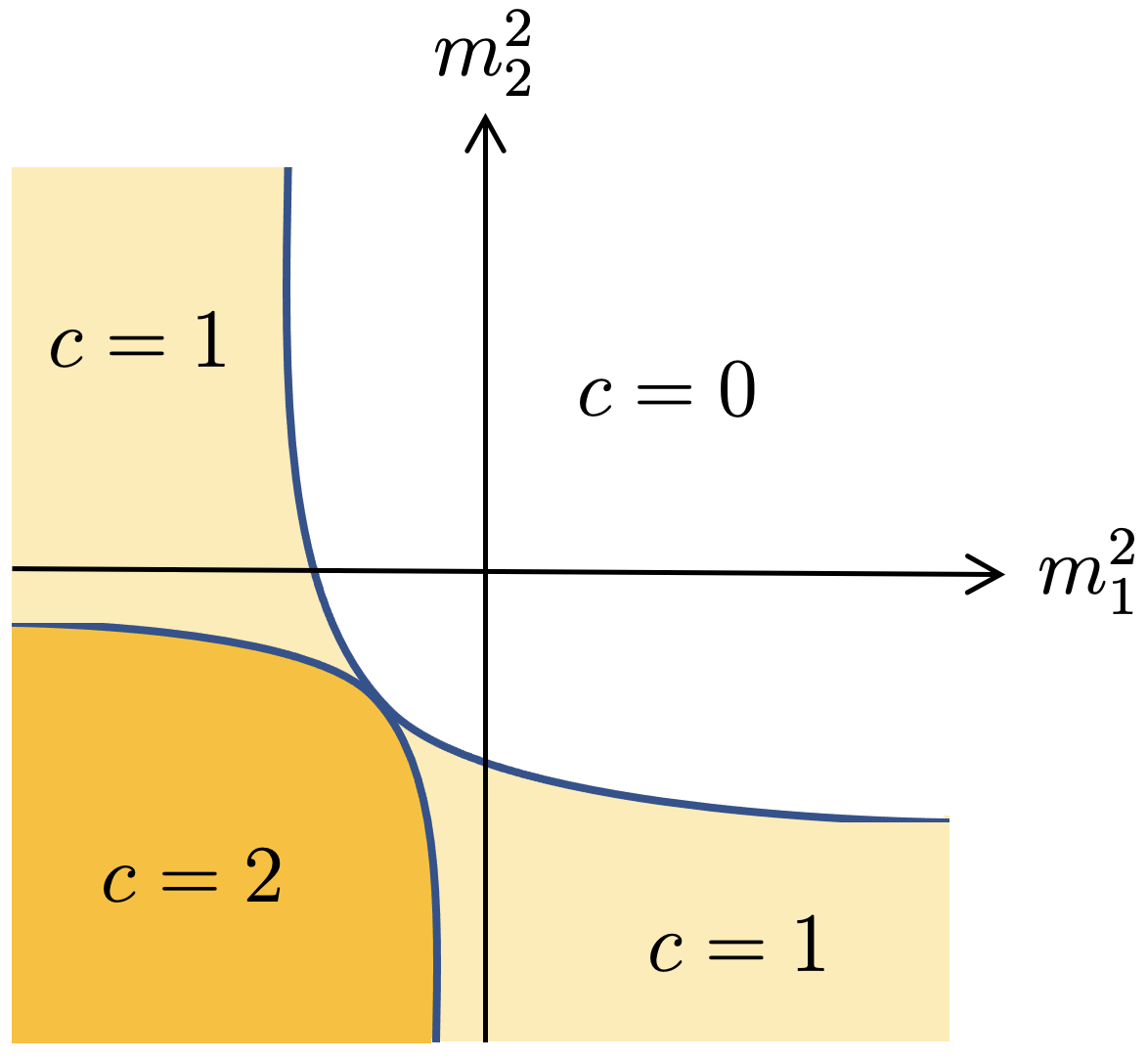}
\captionof{figure}{The phase diagram for the chiral gauge theory for symmetric mass generation.}\label{smgfig}
\end{center}
\noindent
with $\widehat{G}_2$ acting trivially, and $\hat{A}_1$ coupled with charge $1$ to the vector symmetry. We finally use the fermionic T-duality (\ref{eq: fermionic T-duality}), which exchanges vector and axial symmetries, to identify this theory as a $c=1$ fermionic CFT with radius given in (\ref{eq: c=1 radius 2}) with $\hat{A}_1$ coupled with charge $1$ now to the axial current. Once again, we have an exact match with the asymptotic analysis.

\para
We can now come back to our earlier question: how are the two $c=1$ regions separated? We propose that there exists a distinguished point in the phase diagram at which the $c=2$ and $c=0$ regions meet, resulting in the phase diagram in Figure \ref{smgfig}.

\para
To show this, we need to argue that there exists a point on the boundary of the stable region of the $c=2$ moduli space at which, rather than flowing to a $c=1$ theory, the theory instead flows to a trivially gapped phase. Indeed, we previously considered the scenario that $\mathcal{O}_{1,0}$ becomes relevant first, and, separately, that $\mathcal{O}_{0,1}$ becomes relevant first. But with two UV parameters to tune, we can find a point at which they both become relevant simultaneously. At this point, generating then the operator $a\cos(\rho_1) + b\cos(2\tilde{\rho}_2)$, for any non-zero $a,b$, trivially gaps the theory.

\para
We finally note that there is a scenario in which we can make this more precise. The UV Lagrangian for the theory (\ref{eq: SMG theory}) is invariant under
\be
  &&\psi_1 \to \psi_2,\quad 
  \psi_2 \to \psi_1^\dagger,\quad 
  \tilde{\psi}_1\to \tilde{\psi}_2^\dagger,\quad 
  \tilde{\psi}_2\to \tilde{\psi}_1,\quad 
  \nn\\
 && \phi_1 \to \phi_2^\dagger, \quad 
  \phi_2 \to \phi_1, \quad 
  A_1\to - A_2,\quad 
  A_2 \to  A_1 \ ,
\label{eq: C' symm}
\ee
provided that  we also swap gauge couplings $e_1\leftrightarrow e_2$, scalar masses $m_1\leftrightarrow m_2$, and quartic couplings $\lambda_1\leftrightarrow \lambda_2$. We specialise to the case $e_1=e_2$ and $\lambda_1 = \lambda_2$. We first learn that the phase diagram is symmetric about the line $m_1=m_2$. More interestingly, if we sit directly on the line $m_1=m_2$, this now defines a new symmetry of the theory. Indeed, this symmetry squares to charge conjugation, and thus we can think of the theory at $m_1=m_2$ as possessing a novel $\Z_4$ charge conjugation symmetry defined by \eqn{eq: C' symm}.

\para 
Using the bosonisation dictionary, we deduce that this symmetry acts on fields in the Higgs phase Lagrangian (\ref{eq: SMG Higgs phase}) as
\begin{align}
  (\phi_1,\phi_2,\sigma_1,\sigma_2,A_1,A_2) \to (-2\tilde{\phi}_2,2\tilde{\phi}_1,-\sigma_2,\sigma_1,-A_2,A_1) \ ,
\label{eq: C' action}
\end{align}
where $\tilde{\phi}_{1,2}$ denote the dual scalars. Indeed, acting with this transformation and then performing a double T-duality on $\tilde{\phi}_1,\tilde{\phi}_2$, one comes back to the same action.

\para
This new symmetry then acts on the spectrum as
\begin{align}
    \hat{\mathcal{O}}_{\mathbf{n},\mathbf{m}} \to \hat{\mathcal{O}}_{\mathbf{n}',\mathbf{m}'},\qquad \mathbf{n}'=\frac{1}{2}(m_2,-m_1),\,\, \mathbf{m}'=2(n_2,-n_1) \ ,
\end{align}
which one can verify is indeed a symmetry of the spectrum (\ref{eq: c=2 spectrum}) using $v_1=v_2$. (This statement continues to be true for any metric $G$ with ${\rm det}\,G=1/4$.) The operators $\mathcal{O}_{r,s}$ that can be dynamically generated in the Lagrangian are thus mapped as
\begin{align}
    \mathcal{O}_{r,s}\to \mathcal{O}_{s,r} \ .
\end{align}
It follows that the first operator that can be generated as we move diagonally into the phase diagram along $m_1=m_2$ is the $\Z_4$ singlet
\begin{align}
    \mathcal{O}_{1,0} + \mathcal{O}_{0,1} = \cos(\rho_1) + \cos(2\tilde{\rho}_2) \ ,
\end{align}
which does indeed drive the theory to a trivially gapped phase.

\section*{Acknowledgements} We're grateful to Ofer Aharony, Guillermo Arias-Tamargo,  Philip Boyle Smith, Aleksey Cherman, Diego Delmastro, Angelo Esposito, Chris Hull, Max Hutt, and Igor Klebanov for useful comments and discussions. This work was supported by the STFC grant ST/X000664/1, and a Simons Investigator Award.
The work of R.M. was supported by the UK Engineering and Physical Sciences grant EP/Z000106/1.
For the purpose of
open access, the author has applied a Creative Commons Attribution (CC BY) licence to any Author Accepted Manuscript version arising from this submission.

\newpage
\appendix

\section{Appendix: Integrating out gauge fields}\label{app: integrating out gauge fields}

We consider 2d QED with a single compact boson of charge $q\neq 0$, which without loss of generality we can take positive, $q>0$. It is well-known that this theory is gapped, and has $q$ vacua. It will be useful to see how this comes about.
\para 
In Euclidean signature, and in the dual frame for the scalar, we have the action\footnote{In this appendix it will prove convenient to use form notation.}
\begin{align}
  S = \int \left[\frac{1}{2e^2}F\wedge \star F + \frac{R^2}{4\pi}d\varphi \wedge \star d\varphi + \frac{iq}{2\pi}\varphi \, dA\right] \ .
\label{eq: appendix action}
\end{align}
It is easy to see that integrating out $A$ generates a mass for $\varphi$ proportional to $e$. The question we want to answer is: how many vacua does this theory have?

\para
There is a cavalier way to proceed. We can take $e^2\to\infty$ in the above action and thus consider the action
\begin{align}
  S = \int \left[ \frac{R^2}{4\pi}d\varphi \wedge \star d\varphi + \frac{iq}{2\pi}\varphi \, dA\right] \ .
\end{align}
We can then split the gauge field as
\begin{align}
  A = A_n + \tilde{A} \ ,
\label{eq: A split}
\end{align}
where $A_n$ is some configuration that carries $n$ units of magnetic flux,
\begin{align}
  \frac{1}{2\pi}\int dA_n = n \ ,
\end{align}
and is thus $A_n$ is only locally defined. Meanwhile $\tilde{A}$ is globally defined and thus $\int d\tilde{A}=0$. Integrating out $\tilde{A}$ then sets $\varphi=c$, constant, and we are left with action
\begin{align}
  S = \frac{iqc}{2\pi}\int dA_n = iqcn \ .
\end{align}
Now summing over topological sectors $n\in \Z$, the path integral $Z = \int D\varphi \, e^{-S}$ localises to configurations $\varphi = c = 2\pi m/q$ with $m=0,1,\dots,q-1$. The theory thus has $q$ vacua.

\para
Of course, the step of sending $e^2\to\infty$ at the level of the Lagrangian was cheating. So let's do better. We now tackle the full action (\ref{eq: appendix action}), and once again make the split (\ref{eq: A split}). We can furthermore choose $A_n$ such that\footnote{We fix the volume $\int\star 1 = 1 $.} $dA_n = 2\pi n \star 1$ and so in particular $d\star d A=0$. This ensures that the Maxwell term splits cleanly into two pieces, and we find the action in the $n$ sector
\begin{align}
  S_n = \frac{2\pi^2 n^2}{e^2} + \int \left[\frac{1}{2e^2}d\tilde{A}\wedge \star d\tilde{A} + \frac{R^2}{4\pi}d\varphi\wedge \star d\varphi + i q n \varphi \star 1 + \frac{iq}{2\pi}\varphi \, d\tilde{A}\right] \ .
\end{align}
Integrating out $\tilde{A}$ imposes
\begin{align}
  \frac{1}{e^2} d\star d \tilde{A} + \frac{iq}{2\pi}d\varphi =0 \ .
\end{align}
This is solved locally by
\begin{align}
  d\tilde{A} = - \frac{iqe^2}{2\pi}(\varphi - a)\star 1 \ ,
\end{align}
for any constant $a$. But recall $\tilde{A}$ is globally defined, implying $\int d\tilde{A}=0$, and hence $a$ must take the value
\begin{align}
  a = \int \varphi \star 1 \ .
\end{align}
Plugging this back into the action, we find partition function
\begin{align}
  Z 	&= \sum_{n\in \Z}\int D\varphi \,e^{-S_n}	\nn\\
  		&= \int D\varphi \left[\sum_{n\in \Z}\exp\left(-\frac{2\pi^2 n^2}{e^2} - i q n\int \varphi \star 1\right)\right]	\nn\\
  		&\hspace{20mm}\times \exp \left[\frac{q^2 e^2}{8\pi^2}\left(\int \varphi \star 1\right)^2-\int \left(\frac{R^2}{4\pi}d\varphi \wedge \star d\varphi + \frac{q^2e^2}{8\pi^2}\varphi^2 \star 1\right)\right] \ .
\end{align}
At this point we do something a little unusual: we use Poisson resummation to write
\begin{align}
  \sum_{n\in\Z}\exp \left[-\frac{2\pi^2 n^2}{e^2} - i qn \int \varphi \star 1\right] = \frac{e}{\sqrt{2\pi}} \sum_{m\in \Z} \exp\left[- \frac{e^2}{2}\left(m - \frac{q}{2\pi}\int \varphi \star 1\right)^2\right] \ .
\end{align}
Plugging this in, we find
\begin{align}
  Z\sim \sum_{m\in \Z} \int D\varphi \exp \left[- \int \left(\frac{R^2}{4\pi}d\varphi \wedge \star d\varphi + \frac{q^2e^2}{8\pi^2}\left(\varphi - \frac{2\pi m}{q}\right)^2 \star 1\right)\right] \ .
\end{align}
If $q=1$ then the sum over $m$ just ensures that the path integrand is indeed $2\pi$-periodic in $\varphi$. In particular in this case there is a single vacuum $\varphi =0 $ (mod $2\pi$). In general however we see that there is a $q$-fold degeneracy of vacua,
\begin{align}
  \varphi =0,\frac{2\pi}{q},\frac{4\pi}{q}\dots, \frac{2\pi (q-1)}{q} \quad (\text{mod }2\pi) \ ,
\end{align}
which is what we wanted to show.

\bibliographystyle{utphys}
\bibliography{biblio}

\end{document}